\documentclass[]{IEEEtran}
\usepackage{amsmath,amsfonts,amssymb,amsmath}
\usepackage{enumerate}
\usepackage{xcolor,soul,framed} %,caption
\usepackage{mathrsfs,sobolev}

\usepackage[pdftex]{graphicx}

\usepackage{url}

\newcommand{\rb}{\mathbf{r}}

\newcommand{\jt}{\mathbf{j}^\perp}
\newcommand{\jl}{\mathbf{j}^\parallel}
\newcommand{\sinc}[1]{\text{sinc}\left\{#1\right\}}
\newcommand{\n}{\hat{\mathbf{n}}}
\newcommand{\dr}{{\bf r} - {\bf r}'}
\newcommand{\Pn}{- \hat{\mathbf{n}} \times \hat{\mathbf{n}} \times }
\newcommand{\gd}{g_{d\pm}}

\newcommand{\Id}{\text{I}}
\newcommand{\je}{\mathbf{j}_e}
\newcommand{\jm}{\mathbf{j}_m}

\newcommand{\jen}{\text{J}_e}
\newcommand{\jmn}{\text{J}_m}

\newcommand{\Tot}{\boldsymbol{\mathcal{T}}_0^\perp}
\newcommand{\Tol}{\boldsymbol{\mathcal{T}}_0^\parallel}
\newcommand{\Te}{\boldsymbol{\mathcal{T}}_-}
\newcommand{\Ti}{\boldsymbol{\mathcal{T}}_+}
\newcommand{\Ten}{\text{T}_-}
\newcommand{\Tin}{\text{T}_+}
\newcommand{\Tie}{\boldsymbol{\mathcal{T}}_\pm}

\newcommand{\dTie}{\boldsymbol{\mathcal{T}}_{d\pm}}

\newcommand{\Ko}{\boldsymbol{\mathcal{K}}_0}
\newcommand{\Ke}{\boldsymbol{\mathcal{K}}_-}
\newcommand{\Ki}{\boldsymbol{\mathcal{K}}_+}
\newcommand{\Ken}{\text{K}_-}
\newcommand{\Kin}{\text{K}_+}

\newcommand{\dKie}{ \boldsymbol{\mathcal{K}}_{d\pm}}
\newcommand{\Kie}{\boldsymbol{\mathcal{K}}_\pm}

\newcommand{\Ei}{\mathbf{E}_{inc}}
\newcommand{\re}[1]{\text{Re}\left\{#1\right\}}

\newcommand{\eo}{\mathbf{e}_0 }
\newcommand{\ho}{\mathbf{h}_0 }

\newcommand{\eon}{\text{E}_0 }
\newcommand{\hon}{\text{H}_0 }
\newcommand{\Zi}{\zeta^+}
\newcommand{\Ze}{\zeta^-}

\newcommand{\dotprd}[3]{\langle #1 \, | #2 | \, #3 \rangle}  

\newcommand{\rp}{\left( \mathbf{r} \right)}

\begin{document}

\title{Static surface mode expansion \\ for the full-wave scattering from penetrable objects}

\author{\IEEEauthorblockN{Carlo Forestiere, Giovanni Gravina, Giovanni Miano, Guglielmo Rubinacci, Antonello Tamburrino}
\IEEEcompsocitemizethanks{\IEEEcompsocthanksitem
C. Forestiere, G. Gravina G. Miano, G. Rubinacci are with the Department of Electrical Engineering and Information Technology, Universit\`{a} degli Studi di Napoli Federico II, via Claudio 21,  Napoli, 80125, Italy
 \IEEEcompsocthanksitem G. Gravina is with 10th Aircraft Maintenance Department, Italian Air Force, Viale dell'Aeronautica 1, 73013 Galatina, Lecce, Italy
\IEEEcompsocthanksitem A. Tamburrino is with Dipartimento di Ingegneria Elettrica e dell'Informazione  ``M. Scarano", Università degli Studi di Cassino e del Lazio Meridionale, Via G. Di Biasio n. 43, 03043 Cassino (FR), Italy
}}

% ====================================================================
\maketitle

% === ABSTRACT ====================================================================
% =================================================================================
\begin{abstract}
%\boldmath
We introduce the longitudinal and transverse static surface modes and use them to solve the full-wave electromagnetic scattering problem from penetrable objects. The longitudinal static modes are the eigenmodes with zero surface curl of the electrostatic integral operator that gives the  tangential component of the electric field, as a function of the surface charge density. The transverse static modes are the eigenmodes with zero surface divergence of the magnetostatic integral operator that returns the tangential component of the vector potential, as a function of the surface current distribution. The  static  modes  only  depend  on  the  shape  of  the object, thus, the same static basis can be used regardless of the frequency of operation and of the material constituting the object. We expand the unknown surface currents of the Poggio-Miller-Chang-Harrington-Wu-Tsai surface integral equations in terms of the static surface modes and solve them using the Galerkin-projection scheme. The static modes expansion allows the regularization of the singular integral operators and yields a drastic reduction of the number of unknowns compared to a discretization based on sub-domain basis functions. The introduced expansion  significantly reduces the cpu-time required for the numerical solution of the scattering problem from particle arrays.
\end{abstract}

% === KEYWORDS ====================================================================
% =================================================================================

\begin{IEEEkeywords}
Electromagnetic scattering, Eigenvalues and eigenfunctions, Resonance, Resonators, Computational Electromagnetics, Integral Equations, Plasmonics, Dielectric Resonators.
\end{IEEEkeywords}

% === I. INTRODUCTION =============================================================
% =================================================================================

\section{Introduction}
The analysis and design of the electromagnetic scattering from a collection of mutually coupled objects is of great importance for many applications, spanning from antenna arrays \cite{collin_antennas_1985} to metasurfaces \cite{yu_flat_2014} and metalens \cite{khorasaninejad_achromatic_2015}. In this context, the use of integral formulations is appealing, since the unknowns are defined only within the objects' volume or, if the objects are spatially piecewise homogeneous, on their boundary and internal interfaces, while the radiation condition at infinity is naturally satisfied. Nevertheless, the corresponding discrete problem is characterized by dense matrices, and their inversion is usually associated with high computational burden, even when acceleration techniques  like fast multipole algorithms are implemented \cite{chew_fast_2000}.

Accurate and efficient solutions of integral formulations heavily depend on the choice of basis functions. Two macro-categories of basis functions may be identified: {\it sub-domain} functions, which are non-zero only over a portion of the object, or {\it entire-domain} functions, which extend over the entire domain of the object. Although sub-domain functions may have a wider applicability, and are arguably more robust when dealing with objects of irregular shape and sharp corners, entire-domain functions are very appealing when multiple scattering problems are considered, where the electromagnetic system under investigation is a collection of mutually-coupled  objects \cite{bucci_use_1995,angiulli_characteristic_1998,doicu_light_2006}.

Representative examples of sub-domain functions are those involved in the divergence-conforming Galerkin method: for instance the Rao-Wilton-Glisson functions \cite{rao_electromagnetic_1982}, loop/star functions \cite{bossavit_computational_1998,albanese_integral_1988,rubinacci_broadband_2006,wilton_novel_1993}, loop/tree functions \cite{wu_study_1995}, Trintinalia-Ling functions \cite{trintinalia_first_2001}, Buffa-Christiansen functions \cite{buffa_dual_2007}, higher order vector basis functions of Nedelec type \cite{graglia_higher_1997}, etc.

Dually, classic examples of entire-domain basis functions are the vector spherical wave functions (see for instance \cite{bohren_absorption_1998}), and the vector spheroidal wave functions \cite{li_spheroidal_2004}. Analytical entire-domain bases may be generated in coordinate systems where the Helmholtz equation is separable.  A different but effective strategy to generate entire domain basis functions even in irregular domains is to introduce a convenient auxiliary eigenvalue problem. This is done for instance with characteristic modes \cite{garbacz_modal_1965,chang_surface_1977,harrington_characteristic_1972,chen_characteristic_2015}, (see \cite{bucci_use_1995,angiulli_characteristic_1998} for arrays of perfectly conductive particles and \cite{faenzi_metasurface_2019} for perfectly conductive metasurfaces). The characteristic modes do  not depend on the particular excitation conditions, and they are effective in the numerical solution of the problems of electromagnetic scattering from collections of  objects of given material at a fixed operating frequency. Nevertheless, characteristic modes do depend on the frequency, and their interesting properties are lost if they are used as a basis at a frequency different from the one at which they are computed. Thus, they may not be the best choice when multiple frequencies are involved since they have to be recalculated at each frequency.

In this paper, we introduce a different set of entire domain basis functions that we call ``static" surface current modes. These modes are the union of two sets: longitudinal and transverse modes, exhibiting vanishing surface curl and surface divergence, respectively. We assemble these two sets by solving two \textit{auxiliary} frequency-independent eigenvalue problems, involving Hermitian and positive-definite surface integral operators, having the {\it static} Green's function as kernel. The static modes are the low-frequency limit of the resonance modes of surfaces of finite conductivity \cite{forestiere_electromagnetic_2019}. Their volume counterparts have been presented in  \cite{mayergoyz_electrostatic_2005,miano_numerical_2010} and in  \cite{forestiere_magnetoquasistatic_2020} where they have been already used to expand the electromagnetic field \cite{forestiere_quantum_2020,forestiere_operative_2022,forestiere_resonance_2020,forestiere_time-domain_2021}.

The static modes can be also considered as ``high level expansion functions", belonging to the same categories of the functions introduced in  \cite{suter_subdomain_2000,prakash_characteristic_2003,matekovits_analysis_2007,freni_fast-factorization_2011}, which are typically used in the solution of integral equations in electrically large structure to reduce the number of unknowns. They are not bound to the conventional discretization limit of $\lambda/20$, to which the local basis function discretization is constrained. The proposed basis shares similarities with the one introduced by Vecchi et al. and used in a hybrid spectral-spatial method for the analysis of printed antennas \cite{vecchi_hybrid_1996,vecchi_numerical_1997}.

We demonstrate that the use of the static modes in the Galerkin projection of the Poggio-Miller-Chang-Harrington-Wu-Tsai surface integral formulation \cite{wu_scattering_1977,chang_surface_1977,poggio_chapter_1973,harrington_field_1993} leads to: i) a regularization of the scattering integral operator; ii) a drastic reduction of the number of unknowns with respect to sub-domain basis functions without deteriorating the accuracy of the solution; iii) a reduction of the cpu-time required for the solution of multiple scattering problems.

The paper is organized as follows: in Sec. \ref{sec:StaticBasis} we introduce the static basis;  in Sec. \ref{sec:PMCWHT} we recall the Poggio-Miller-Chang-Harrington-Wu-Tsai formulation and we show that the static mode expansion regularizes the involved operator, and that a proper rescaling of the unknowns makes this formulation immune from the low-frequency breakdown. In Sec. \ref{sec:Results}, we validate the introduced method in two resonant scattering problems, namely the scattering from a metal particle and high-permittivity particle in the visible/near-infrared spectral range. Eventually, we apply this method to the solution of a multiple scattering problem. In Sec. \ref{sec:Conclusion} we draw the conclusions.

\section{Static surface current modes}
\label{sec:StaticBasis}

We denote with $\Omega$ a bounded three-dimensional domain, whose boundary $\partial \Omega$ is ``sufficiently regular" \cite{monk_finite_2003}; $\n$ is the normal to $\partial \Omega$ pointing outward. A sufficiently smooth vector field $\mathbf{j}$ defined on a regular surface $\partial \Omega$ can be resolved into the sum of two components \cite{scharstein_helmholtz_1991,nair_generalized_2011}: an irrotational and non-solenoidal vector field $\mathbf{j}^\parallel$ and a solenoidal and rotational (non zero curl) vector field $\mathbf{j}^\perp$.
The vector fields $\mathbf{j}^\parallel$ and $\mathbf{j}^\perp$ are orthogonal according to the scalar product
\begin{equation}
    \langle {\bf C}, {\bf D} \rangle = \int_{\partial \Omega} {\bf C}^* \rp \cdot \mathbf{D} \rp dS.
    \label{eq:ScaPro}
\end{equation}
In the following, we introduce a basis for each of the two components.

\subsection{Longitudinal static modes}
The longitudinal static surface current modes (called in the following {\it longitudinal static modes}  for brevity) are nontrivial solutions of the eigenvalue problem:
\begin{equation}
     \Tol \{ \jl_k \} \left( {\bf r} \right)  =  \gamma_k^\parallel  \, \jl_k \qquad \text{on} \; \partial \Omega,
    \label{eq:AuxProb_Tl}
\end{equation}
where
\begin{equation}
    \Tol \left\{ {\bf w} \right\} \left( {\bf r} \right) = \n \times \n \times \nabla \oint_{\partial\Omega} g_0 \left( {\bf r}  - {\bf r'} \right) \nabla' _S  \cdot {\bf w}\left( {\bf r}' \right) dS', 
    \label{eq:Operator_Tlo}
\end{equation}
and $g_0$ is the homogeneous space static Green's function
\begin{equation}
  g_0 \left( {\bf r} - {\bf r}' \right) = \frac{1}{4 \pi} \frac{ 1}{ \left|{\bf r} - {\bf r}' \right| }.
\end{equation}
Apart from a multiplicative factor, the integral operator \ref{eq:Operator_Tlo} gives the tangential component of the static electric field generated by a surface charge density distribution. Its spectrum has the following properties (see \cite{mayergoyz_electrostatic_2005}):

\begin{enumerate}[i]
    \item the eigenvalues $\{ \gamma_k^\parallel \}$ and the corresponding eigenmodes $\{ \jl_k \}$ depend on the shape of the object, but are independent of the object material, and of the frequency of operation;
    \item the eigenvalues are real and positive;
    \item the eigenmodes are orthonormal according to the scalar product \ref{eq:ScaPro};
    \end{enumerate}
The eigenvalues of a spherical surface of unit radius have the analytical expression
        \begin{equation}
        {\gamma_n^\parallel} = \frac{n \left( n + 1 \right)}{\left(2 n + 1\right)} \qquad n=1,2,3\ldots.
        \label{eq:EigLongSphere}
    \end{equation}
        We now introduce a spherical coordinate system. The spherical coordinates of the point with position vector $\rb$ are $(r,\theta,\phi)$ (with $0 \le r < \infty$, $0\le \theta < \pi$ and  $0\le \phi<2\pi$). The basis for the three-dimensional vector space is the set $(\hat{\rb}, \hat{\boldsymbol{\theta}},\hat{\boldsymbol{\phi}})$, where $\hat{\rb}$ is the radial unit vector, $\hat{\boldsymbol{\theta}}$ is the polar unit vector, and $\hat{\boldsymbol{\phi}}$ is the azimuthal unit vector.  The eigenmodes corresponding to $\gamma_n^\parallel$ are the vector spherical harmonics $\mathbf{W}_n^m(\theta,\phi)$ where $m$ is an integer such that $-n \le m \le n$: 
    \begin{equation}
    \mathbf{j}_{mn}^\parallel = \mathbf{W}_n^m = \hat\rb\times \mathbf{X}_n^m=\frac{1}{\sqrt{n(n+1)}} \nabla Y_n^m.
    \end{equation}
    The spherical harmonic $Y_n^m \left( \theta,\phi\right)$ of degree $n$ and order $m$ is given by
\begin{equation}
     Y_n^m(\theta,\phi)=C_{mn} P_n^{|m|}(\cos\theta)e^{im\phi}
\end{equation}
where $P_n^m(\cos\theta)$ is the associated Legendre polynomial of degree $n$ and order $m$, and $C_{mn}$ is a normalization coefficient. The spherical harmonics are orthogonal. We normalize them in such a way that
\begin{equation}
    \int_{}\,|Y_n^m(\theta,\phi)|^2\,d\Omega = 1,
\end{equation}
where $\int_{}(\cdot)d\Omega = \int_{0}^{\pi}d\theta\, \sin\theta\int_{0}^{2\pi}d\phi\,(\cdot)$. The normalization coefficient $C_{mn}$ is equal to
\begin{equation}
    C_{mn} = \sqrt{\frac{2n+1}{4\pi}\frac{(n-m)!}{(n+m)!}}.
\end{equation}
The degree $n$ also determines the multipolar order of the vector spherical harmonics (e.g. $n=1$ for a dipole, $n=2$ for a quadrupole, etc.). Thus, Eq. \ref{eq:EigLongSphere} shows that longitudinal modes associated with larger 
eigenvalues $\gamma^\parallel_n$ are characterized by higher multipolar order.

%In the following, we employ the set  $\{ \jl_k \}$ for the representation of square integrable solenoidal vector fields defined on $\partial \Omega$ with zero surface curl.

\subsection{Transverse static modes}
The transverse static surface current modes (called {\it transverse static modes} in the following for brevity) are non-trivial solutions of the eigenvalue problem:
\begin{equation}
    \label{eq:AuxProb_Tt}
    \Tot \left\{ \jt_k \right\} \left( {\bf r} \right)  = \gamma_k^\perp  \jt_k \qquad \text{on} \; \partial \Omega,
\end{equation}
with 
\begin{equation}
    \label{eq:Operator_Tto}
    \Tot \left\{ {\bf w} \right\} \left( {\bf r} \right) =  \Pn \oint_{\partial \Omega} g_0 \left( \dr \right) {\bf w} \left( {\bf r}' \right) d {S}'.
\end{equation} 
Apart from a multiplicative factor, the  integral operator \ref{eq:Operator_Tto} gives the static vector potential generated by a surface current distribution. Its spectrum has the following properties, that can be derived using the standard methods of eigenvalue problems, analogously to \cite{forestiere_magnetoquasistatic_2020,tamburrino_monotonicity_2021}:

\begin{enumerate}[i]
    \item The eigenvalues $\{ \gamma_k^\perp \}$ and the modes $\{ \jt_k \}$ depend on the shape of the object, and are independent of the object material and of the frequency of operation
    \item the eigenvalues are real and positive;
    \item the modes $\{ \jt_k \}$ associated are orthonormal according to the scalar product \ref{eq:ScaPro}.
\end{enumerate}
The eigenvalues of a spherical surface of unit radius  have the analytical expression
    \begin{equation}
        {\gamma_n^\perp} =  \frac{1}{\left(2 n + 1\right)} \qquad n=1,2,3\ldots.
        \label{eq:EigTransSphere}
    \end{equation}

    The modes corresponding to $\gamma_n^\perp$  are the vector spherical harmonics $\mathbf{X}_n^m(\theta,\phi)$ where $m$ is an integer such that $-n \le m \le n$:
    \begin{equation}
        \mathbf{j}_{mn}^\perp =  \mathbf{X}_{n}^m = \frac{1}{\sqrt{n(n+1)}}\nabla Y_n^m \times \rb.
    \end{equation}
The degree $n$ also determines the multipolar order of the vector spherical harmonics (e.g. $n=1$ for a dipole, $n=2$ for a quadrupole, etc.). Thus, Eq. \ref{eq:EigTransSphere} shows that transverse modes associated with smaller eigenvalues $\gamma^\perp_n$ are characterized by higher multipolar order.

%Any longitudinal static modes is orthogonal to any transverse modes according to the scalar product $\langle {\bf C}, {\bf D} \rangle$. 

\subsection{Computation of the static modes}
Let us introduce a surface triangulation of $\partial \Omega$, with $N_p$ vertices,  $N_t$ elements, and $N_e$ edges. We represent the static modes in term of convenient sub-domain basis functions, namely the loop and star functions \cite{wilton_novel_1993,burton_study_1995}. Specifically, any longitudinal mode $\jl_h$ is expanded in terms of  (non-solenoidal) star basis functions $\left\{ \mathbf{j}_p^{\star} \right\}$ with coefficients $\alpha^{\parallel \, \star}_{h,p}$. Dually, any transverse mode $\jt_h$ is expanded in terms of (solenoidal) loop basis functions $ \left\{ \mathbf{j}_q^\circlearrowleft  \right\}$ with coefficients $\alpha^{\perp  \, \circlearrowleft}_{h,q}$:

\begin{equation}
   \jl_h  = \sum_{p=1}^{N_t-1} \alpha^{\parallel \, \star}_{h,p} \ \mathbf{j}_p^{\star}, \qquad
    \jt_h  = \sum_{q=1}^{N_p-1} \alpha^{\perp  \, \circlearrowleft}_{h,q} \ \mathbf{j}_q^\circlearrowleft.
\end{equation}
The loop functions are divergence free, thus they correctly represent the transverse static modes. Instead, the star functions are not curl free (they are often denoted as quasi-curl \cite{vecchi_hybrid_1996}) thus they only approximately represent the longitudinal static modes. Both star and loop functions admit a linear representation in terms of RWG basis functions \cite{rao_electromagnetic_1982}. 

For closed surfaces with no handles, the number of linearly independent loop functions is $N_p-1$, while the number of linearly independent star functions is $N_t-1$ . Thus, the numerical auxiliary eigenvalue problem for longitudinal static modes is
\begin{equation}
    \text{T}_0^{\parallel {\star \star}} \, \mathbf{J}_h^{\star} = \gamma_h^\parallel \, \text{R}^{{\star \star}} \, \mathbf{J}_h^{\star},
    \label{eq:NumericLongProblem}
\end{equation}
where $(  \text{T}_0^{\parallel \, {\star \star}} )_{pq} = \langle  \mathbf{j}_p^{\star}, \Tol \, \mathbf{j}_q^{\star} \rangle$, 
$\left(  \text{R}^{{\star \star}} \right)_{pq} = \langle  \mathbf{j}_p^{\star}, \, \mathbf{j}_q^{\star} \rangle,
$
and $\mathbf{J}_h^{\star} = [ \alpha^\parallel_{h,1}, \alpha^\parallel_{h,2}, \ldots, \alpha^\parallel_{h,N_t-1} ]^\intercal$. The numerical auxiliary eigenvalue problem for transverse static modes is
\begin{equation}
    \text{T}^{\perp \, \circlearrowleft \circlearrowleft} \, \mathbf{J}_h^{\circlearrowleft} = \gamma_h^\perp \text{R}^{{\circlearrowleft \circlearrowleft}}  \mathbf{J}^{\circlearrowleft},
    \label{eq:NumericTransProblem}
\end{equation}
where 
$ \left(  \text{T}^{\perp \, \circlearrowleft \circlearrowleft} \right)_{pq} = \langle  \mathbf{j}_p^{\circlearrowleft}, \Tol \, \mathbf{j}_q^{\circlearrowleft} \rangle,
    \quad
    \left(  \text{R}^{{\circlearrowleft \circlearrowleft}} \right)_{pq} = \langle  \mathbf{j}_p^{\circlearrowleft}, \, \mathbf{j}_q^{\circlearrowleft} \rangle$,
 and $\mathbf{J}_h^{\circlearrowleft} = [ \alpha^\perp_{h,1}, \alpha^\perp_{h,2}, \ldots,  \alpha^\perp_{h,N_p-1} ]^\intercal$.
 
Since the loop and star functions are not orthogonal,  the matrices $\text{R}^{{\star \star}}$ and $\text{R}^{{\circlearrowleft \circlearrowleft}}$ are not identity matrices, thus Eqs. \ref{eq:NumericLongProblem} and \ref{eq:NumericTransProblem} are \textit{generalized} eigenvalue problems. The involved matrices are real, symmetric, and positive definite. Thus, efficient numerical algorithms for the eigenvalue calculation do apply, such as the Cholesky factorization \cite{golub_matrix_1983}. Moreover, the matrices properties also determine the orthogonality, at the discrete level, of any pair of longitudinal modes, and any pair of transverse modes. The numerical integration of shape functions times the Green’s functions or its gradient are evaluated using the techniques introduced by Graglia \cite{graglia_numerical_1993}.
      
\section{Poggio-Miller-Chang-Harrington-Wu-Tsai (\texttt{PMCHWT}) surface integral equation}
\label{sec:PMCWHT}
A linear, homogeneous, isotropic material occupies the three-dimensional domain $\Omega$.  The material has  permittivity $\varepsilon^+ \left( \omega \right)$, permeability $\mu^+ \left( \omega \right)$ and it is surrounded by a background medium with permittivity $\varepsilon^- \left( \omega \right)$ and permeability $\mu^- \left( \omega \right)$. The object is illuminated by a time harmonic electromagnetic field $\re{\Ei (\mathbf{r}) \, e^{i \omega t}}$.
The equivalent electric $\je$ and magnetic $\jm$ surface current densities, defined on $\partial \Omega$, are solutions of the following surface integral problem formulated by Poggio-Miller-Chang-Harrington-Wu-Tsai (\texttt{PMCHWT}) \cite{wu_scattering_1977,chang_surface_1977,poggio_chapter_1973}:

\begin{equation}
 \boldsymbol{\mathcal{Z}} \,  \mathbf{J} = \mathbf{F},
  \label{eq:PMCHWT}
\end{equation}
where

\begin{equation}
  \boldsymbol{\mathcal{Z}}  = \left( {\begin{array}{*{20}c}
    \Ze \Te +  \Zi \Ti &  \Ke +\Ki
  \\
      -  \left( \Ke +\Ki \right) &  
\Te/\Ze  + \Ti/\Zi  \\
\end{array}} \right),
  \label{eq:PMCHWTbis}
\end{equation}
\begin{equation}
  \mathbf{J} =  \left[ \je,  \jm \right]^\intercal, \qquad
     \mathbf{F} = \left[     \eo,  \ho  \right]^\intercal,
\end{equation}
\begin{equation}
    \mathbf{e}_0  =  \left. \Pn {\bf{E}}_{inc}  \right|_{\partial \Omega}, \quad
     \mathbf{h}_0  =  \left. \Pn {\bf{H}}_{inc}  \right|_{\partial \Omega}.
\end{equation}
The operators $\Kie$ and $\Tie$ are the \texttt{MFIE} and \texttt{EFIE} integral operators: 
\begin{subequations}
  \begin{align}
    \label{eq:Operator_K}
    \Kie & \left\{ {\bf w} \right\} \left( {\bf r} \right) = \n \times \n \times \int_{\partial \Omega} {\bf w}  \left( {\bf r}' \right) \times \nabla' g^\pm \left( {\bf
r}  - {\bf r}' \right) dS',  \\
    \label{eq:Operator_T}
    \notag
    \Tie & \left\{ {\bf w} \right\} \left( {\bf r} \right) =    j k^\pm \n \times \n \times  \int_{\partial \Omega} g^\pm \left( \dr \right) {\bf w} \left( {\bf r}' \right) dS'\\ & + \frac{1}{j k^\pm} \n \times \n \times  \int_{\partial \Omega} \nabla' g^\pm \left( \dr \right) \nabla' _S  \cdot {\bf w}\left( {\bf r}' \right) dS',
    \end{align}
\end{subequations}
$\nabla_\text{S} \cdot$ denotes the surface divergence, $g^\pm$ is the homogeneous space Green's function of the region $\Omega_\pm$, i.e.
\begin{equation}
  g^\pm \left( {\bf r} - {\bf r}' \right) = \frac{ e^{- j k^\pm \left|{\bf r} - {\bf r}' \right|} }{4 \pi \left|{\bf r} - {\bf r}' \right| },
  \label{eq:GreenFunctin}
\end{equation}
$k^\pm =\omega \sqrt{\mu^\pm \varepsilon^\pm}$, and $\zeta^\pm = \sqrt{{\mu^\pm}/{\varepsilon^\pm}}$.

\subsection{Galerkin equations}

Aiming at the solution of the \texttt{PMCWHT} equation \ref{eq:PMCHWT},  we represent the equivalent electric and magnetic surface currents in terms of the transverse static modes $\{ \jt_p \}_{p = 1 \ldots \text{N}^\perp}$ associated with the first $\text{N}^\perp$ eigenvalues $\gamma_p^\perp$ (sorted in descending order), and in terms of the longitudinal static modes $\{ \jl_q \}_{q = 1 \ldots \text{N}^\parallel}$  associated with the first  $\text{N}^\parallel$ eigenvalues $\gamma_q^\parallel$ (sorted in ascending order), namely
\begin{equation}
\left\{ 
   \begin{aligned}
      \je \rp \approx \sum_{p = 1}^{\text{N}^\perp} {\alpha_p^\perp } \, \jt_p \rp + \sum_{q = 1}^{\text{N}^\parallel} {\alpha_q^\parallel } \, \jl_q \rp,  \\
      \jm \rp \approx \sum_{p = 1}^{\text{N}^\perp} {\beta_p^\perp } \, \jt_p \rp + \sum_{q = 1}^{\text{N}^\parallel} {\beta_q^\parallel } \, \jl_q \rp.
  \end{aligned}
\right.
\label{eq:SurfaceCurrentExpansion}
\end{equation}
The choice of sorting the longitudinal eigenvalues accordingly to an ascending order and the transverse eigenvalues accordingly to a descending order guarantees that low-index eigenvalues are associated with modes of low-order multipolar order (dipole, quadrupole, octupole ...).

Therefore, we define the unknown block vectors
\begin{equation}
    \text{J}_e = \left( \boldsymbol{\alpha}^\perp | \boldsymbol{\alpha}^\parallel \right)^\text{T}, \qquad \text{J}_m = \left( \boldsymbol{\beta}^\perp | \boldsymbol{\beta}^\parallel \right)^\text{T},
\end{equation}
with $\boldsymbol{\alpha}^a = \left[\alpha_1^a,\alpha_2^a,\ldots,\alpha_{\text{N}^a}^a \right]^\intercal $ and with $\boldsymbol{\beta}^a = \left[\beta_1^a,\beta_2^a,\ldots,\beta_{\text{N}^a}^a \right]^\intercal $ and $a = \parallel, \perp $

We find the finite dimensional approximation of the \texttt{PMCHWT} problem by substituting Eq. \ref{eq:SurfaceCurrentExpansion} in Eq. \ref{eq:PMCHWT} and by projecting along the same set of modes, accordingly to a Galerkin projection scheme:
\begin{equation}
\text{Z} \, \text{J} =
\left(
\begin{array}{c}
    \eon   \\
    \hon  
\end{array}
\right),
  \label{eq:PMCHWTn}
\end{equation}
where
\begin{equation}
    \text{J} = 
    \left(
\begin{array}{c}
    \jen   \\
    \jmn  
\end{array} \right),
\end{equation}
\begin{equation}
    \text{Z} =  \left( {\begin{array}{*{20}c}
    \Ze \Ten +  \Zi \Tin &  \Ken +\Kin
  \\
      -  \left( \Ken +\Kin \right) &  
\Ten/\Ze  + \Tin/\Zi  \\
\end{array}} \right),
\end{equation}
\begin{equation}
    \text{T}_\pm = \left( 
    \begin{array}{c|c}
      \text{T}_\pm^{\perp,\perp}   & \text{T}_\pm^{\perp,\parallel}  \\
      \hline
        \text{T}_\pm^{\parallel,\perp}  & \text{T}_\pm^{\parallel,\parallel} 
    \end{array} \right), \qquad
    \text{K}_\pm = \left( 
    \begin{array}{c|c}
      \text{K}_\pm^{\perp,\perp}   & \text{K}_\pm^{\perp,\parallel}  \\
      \hline
        \text{K}_\pm^{\parallel,\perp}  & \text{K}_\pm^{\parallel,\parallel} 
    \end{array} \right),
\end{equation}

\begin{equation}
  \left( \text{K}_\pm^{a\,b} \right)_{pq} =  \dotprd{\mathbf{j}^a_p}{\Kie}{\mathbf{j}^b_q}, \quad
  \left( \text{T}_\pm^{a\,b} \right)_{pq} =  \dotprd{\mathbf{j}^a_p}{\Tie}{\mathbf{j}^b_q },
\end{equation}
\begin{equation}
    \text{E}_0 = ( \text{E}_0^\perp | \text{E}_0^\parallel )^\text{T}, \qquad \text{H}_0 = ( \text{H}_0^\perp | \, \text{H}_0^\parallel )^\text{T},
\end{equation}
and
\begin{equation}
        ( \text{E}_0^a  )_p = \langle \mathbf{j}_p^a, \eo \rangle 
    \qquad
    \left( \text{H}_0^a  \right)_p = \langle \mathbf{j}_p^a, \ho \rangle,
\end{equation}
with $a,b = \parallel,\perp$. The finite dimensional system has $2 ( \text{N}^\parallel + \text{N}^\perp )$ degrees of freedom.

We now decompose the Green's function as the sum of the static Green's function $g_0$ and a regular difference term $\gd $:
\begin{equation}
  g_\pm \left( \dr \right) = g_0 \left( \dr \right)
  +  \gd \left( \dr \right),
  \label{eq:greenSum}
\end{equation}
where
\begin{equation}
  \gd \left( \dr \right)    =   \frac{e^{- j \frac{k^\pm}{2} \left| \dr \right|}}{4 \pi \, j} k_\pm \sinc{ \frac{k^\pm}{2} \left|{\bf r} - {\bf r}' \right|}.
\end{equation}
In the past, the splitting of the Green’s function into its static component and a difference term has been used to introduce well-conditioned
and accurate scheme for the low-frequency analysis of PEC targets with the MFIE \cite{ubeda_divergence-conforming_2011}.
By applying this decomposition to the operators $\Tie$ and $\Kie$, defined in \ref{eq:Operator_K} and \ref{eq:Operator_T}, we obtain:
\begin{equation}
    \begin{aligned}
      \Tie &= +\frac{1}{j k^\pm} \Tol - j k^\pm \Tot + \dTie, \\
      \Kie &= \Ko + \dKie,
    \end{aligned}
\end{equation}
where $\Tol$ and $\Tot$ are the static operators defined in Eqs. \ref{eq:Operator_Tlo} and \ref{eq:Operator_Tto}, and
\begin{multline}
    \dTie \left\{ {\bf w} \right\} \left( {\bf r} \right) =  j k^\pm \, \n \times \n \times  \int_{\partial \Omega} \gd \left( {\bf r}  - {\bf r}' \right) {\bf w} \left( {\bf r}'
\right) dS'   \\   + \frac{1}{j k^\pm} \n \times \n \times  \int_{\partial \Omega} \nabla \gd \left( {\bf r}  - {\bf r'} \right) \nabla' _S  \cdot {\bf w}\left( {\bf r}' \right) dS',
\end{multline}
\begin{equation}
    \Ko \left\{ {\bf w} \right\} \left( {\bf r} \right) = \n \times \n \times \int_{\partial \Omega} {\bf w}  \left( {\bf r}' \right) \times \nabla' g_0 \left( {\bf
r}  - {\bf r}' \right) dS',    
\end{equation}
\begin{equation}
    \dKie \left\{ {\bf w} \right\} \left( {\bf r} \right) = \n \times \n \times \int_{\partial \Omega} {\bf w}  \left( {\bf r}' \right) \times \nabla'  g_{d\pm} \left( {\bf
r}  - {\bf r}' \right) dS'.  
\end{equation}
The above decomposition considerably simplifies the calculation of the finite dimensional operators $\text{T}_\pm^{a\,b}$ with $a,b=\parallel,\perp$, which are obtained by projecting the operator $\boldsymbol{\mathcal{T}}_\pm$ along the longitudinal and transverse static modes because $\{ \jl_k \}$ and $\{ \jt_k \}$ diagonalize the static operators $\Tot$ and $\Tol$:
\begin{subequations}
      \begin{align}
      \notag
\left( \text{T}_\pm^{\parallel\,\parallel} \right)_{pq} &= \langle \jl_p , \Tie \, \jl_q  \rangle = \\ & \frac{ \gamma_p^\parallel }{j k^\pm} \, \delta_{p,q} - j k^\pm \langle \jl_p, \Tot   \jl_q   \rangle + \langle \jl_p, \dTie   \jl_q   \rangle, \\
\left( \text{T}_\pm^{\perp\,\parallel} \right)_{pq} &=         \langle \jt_p , \Tie \, \jl_q \rangle =    \langle \jt_p,
    \dTie   \jl_q  \rangle, \\
\left( \text{T}_\pm^{\parallel\,\perp} \right)_{pq} &=       \langle \jl_p , \Tie \, \jt_q   \rangle =  \langle \jl_p,
    \dTie \,  \jt_q \rangle, \\
  \left( \text{T}_\pm^{\perp\,\perp} \right)_{pq} &=  \langle \jt_p , \Tie \, \jt_q   \rangle = - j k^\pm \gamma_h^\perp \delta_{p,q} +  \langle \jt_p,
    \dTie \,  \jt_q  \rangle,
    \end{align}
    \label{eq:Regularization}
\end{subequations}
where $\delta_{p,q}$ is the Kronecker delta.  We point out that the numerical computation of the terms $\langle \jl_p,  \dTie \,  \jt_q \rangle$ is straightforward, since their kernels are regular functions. These terms are the only ones depending on the operating frequency. 
The  decomposition of the Green's function into the sum of a static term and of a  regular difference term relieves us from the task of computing almost all the integrals with (integrable) singularity, which usually results in longer computational time compared to their regular counterpart. There are however two exceptions: $\langle \jl_p, \boldsymbol{\mathcal{T}}_0^\perp \, \jl_q \rangle $ and $\langle \mathbf{j}^a_p, \boldsymbol{\mathcal{K}}_0^\perp \, \mathbf{j}_q^b \rangle $, $\forall a,b \in \perp,\parallel$. These terms are frequency-independent, thus when the calculation of the scattering response of an object for multiple frequencies of the exciting field is required, they can be conveniently precalculated and stored away, while only the regular terms must be calculated at any frequencies.

\subsection{Low-frequency analysis}
\label{sec:LowFreq}
Surface integral formulations may suffer from ill-conditioning due to low frequency breakdown. The low frequency breakdown phenomenon manifests when the operating wavelength is much larger than the dimension of the object \cite{wilton_improving_1981}, and originates from the different frequency-scaling of the terms associated with the vector and the scalar potentials. This is a common scenario which may be encountered in several applications, including metamaterials and electromagnetic bandgap (EBG) structures, or in the analysis of interconnects and packaging. This problem has been addressed by using quasi-Helmholtz decompositions, such as loop/star \cite{wilton_improving_1981,burton_study_1995,vecchi_loop-star_1999}, loop-tree \cite{wu_study_1995,andriulli_loop-star_2012}, or null-pinv \cite{miano_surface_2005} decomposition, followed by a basis rearrangement \cite{zhao_integral_2000,chen_analysis_2001}. It is also worth to point out that some formulations are immune from this problem, such as the N-Müller formulation \cite{yla-oijala_surface_2018}, or the formulation obtained  by augmenting the traditional EFIE by including charge as extra unknown \cite{taskinen_current_2006,qian_augmented_2008}.

In this section, we summarize the behaviour of the \texttt{PMCHWT} at very low frequencies. In this limit, the frequency dependence of the elements of the matrix $\text{Z}$ follows different scaling laws, which are easily determined by following  \cite{chen_analysis_2001,forestiere_frequency_2017}
\begin{equation}
    \left(
    \begin{array}{cc}
    \text{T}_\pm^{\perp\,\perp} &    \text{T}_\pm^{\parallel\,\perp}  \\
    \text{T}_\pm^{\perp\,\parallel} &      \text{T}_\pm^{\parallel\,\parallel}
    \end{array}
    \right) 
    \xrightarrow{\omega \downarrow 0}
    \left(
    \begin{array}{cc}
    -{j k^\pm} {\Gamma}^\perp &
    j \left( k^\pm \right)^3 \text{T}_2^{\parallel \perp}
    \\
    j \left( k^\pm \right)^3 \text{T}_2^{\parallel \perp}
 &   + {\Gamma}^\parallel / {j k^\pm}
    \end{array} \right),
    \label{eq:LowFrequencyOperatorsT}
\end{equation}
\begin{equation}
    \left(
    \begin{array}{cc}
       \text{K}_\pm^{\perp \perp}  & \text{K}_\pm^{\perp \parallel}  \\
        \text{K}_\pm^{\parallel \perp} & \text{K}_\pm^{\parallel \parallel}
    \end{array}
    \right) \xrightarrow{\omega \downarrow 0} 
    \left(
    \begin{array}{cc}
       \left(k^\pm\right)^2 \text{K}_2^{\perp \perp}  & \text{K}_0^{\parallel \perp }  \\
        \text{K}_0^{\perp \parallel } &  \text{K}_0^{\parallel \parallel}
    \end{array}
    \right),
        \label{eq:LowFrequencyOperatorsK}
\end{equation}
where ${\Gamma}^\perp$ and ${\Gamma}^\parallel$ are diagonal matrices,
$
    {\Gamma}^\perp = \text{diag} \left\{ \gamma^\perp_1,\gamma^\perp_2, \ldots, \gamma^\perp_{\text{N}^\perp} \right\},$
        ${\Gamma}^\parallel = \text{diag} \left\{ \gamma^\parallel_1,\gamma^\parallel_2, \ldots, \gamma^\parallel_{\text{N}^\parallel} \right\},
$
and
\begin{align}
{T}_2^{\parallel \perp} &=  \frac{1}{8 \pi} \int_{\partial \Omega} \jt_p \left( \mathbf{r}  \right) \cdot  \int_{\partial \Omega} { \left| \mathbf{r} - \mathbf{r}' \right|} \, \jt_q \left( \mathbf{r}' \right) dS dS', \\
{K}_2^{\perp \perp} &= -\frac{1}{8 \pi} \int_{\partial \Omega} \jt_p \left( \mathbf{r}  \right) \cdot \int_{\partial \Omega} \frac{ \left( \mathbf{r} - \mathbf{r}' \right) }{  \left| \mathbf{r} - \mathbf{r}' \right|} \times \jt_q \left( \mathbf{r}' \right) dS dS'.
\end{align}
%
%The derivation of the asymptotic expression of the operators projections \ref{eq:LowFrequencyOperatorsT} and  \ref{eq:LowFrequencyOperatorsK}  is straightforward except for ${K}_\pm^{\perp \perp}$. In this case, as originally pointed out in  \cite{chen_analysis_2001}, it is exploited that the magnetostatic field produced by a transverse function is curl-free.
Thus, by using Eqs. \ref{eq:LowFrequencyOperatorsT} and \ref{eq:LowFrequencyOperatorsK} it easy to prove that, in the static limit, the discrete matrix $\text{Z}$ approaches the matrix $\text{Z}_0$ which exhibits the following frequency dependence:
\begin{equation}
    {Z}_0 \propto
    \left[
    \begin{array}{cccc}
     \mathcal{O} \left( \omega \right)  &  \mathcal{O} \left( \omega^3 \right)  &  \mathcal{O} \left( \omega^2 \right)  &  \mathcal{O} \left( 1 \right)   \\
     \mathcal{O} \left( \omega^3 \right)  &  \mathcal{O} \left( \omega^{-1} \right)  &  \mathcal{O} \left( 1 \right)  &  \mathcal{O} \left( 1 \right)   \\
     \mathcal{O} \left( \omega^2 \right)  &  \mathcal{O} \left( 1 \right)  &  \mathcal{O} \left( \omega \right)  &  \mathcal{O} \left( \omega^3 \right)  \\
     \mathcal{O} \left( 1 \right)  &  \mathcal{O} \left( 1 \right)  &  \mathcal{O} \left( \omega^3 \right)  &  \mathcal{O} \left( \omega^{-1} \right)  \\
    \end{array} 
    \right]
\end{equation}
The excitation vector associated with a plane wave exhibits the following dependencies \cite{chen_analysis_2001}:
\begin{equation}
\left[
E_0^\perp,   \\
E_0^\parallel,  \\
H_0^\perp,   \\
H_0^\parallel 
\right]^\intercal \propto
\left[
 \mathcal{O}  \left( \omega \right),
 \mathcal{O}  \left( 1 \right),
\mathcal{O}  \left( \omega \right),
\mathcal{O}  \left( 1 \right)
\right]^\intercal.
\end{equation}
Following \cite{chen_analysis_2001}, we introduce the rearrangement and scaling of the basis
\begin{equation}
    \tilde{\text{Z}} = \text{D}_1 \,   \text{Z}  \, \text{D}_2,
\end{equation}
where
$ {D}_1 = \text{diag} \left\{ k_-^{-1} \Id_{\text{N}^\perp}, \Id_{\text{N}^\parallel}, k_-^{-1} \Id_{\text{N}^\perp},  \Id_{\text{N}^\parallel} \right\}$, ${D}_2 = \text{diag} \left\{  \Id_{\text{N}^\perp}, i k_- \Id_{\text{N}^\parallel},  \Id_{\text{N}^\perp},  i k_- \Id_{\text{N}^\parallel} \right\},$
$\Id_{\text{N}^\perp}$ is the the $\text{N}^\perp \times \text{N}^\perp$ identity matrix, and $\Id_{\text{N}^\parallel}$ is the $\text{N}^\parallel \times \text{N}^\parallel$ identity matrix. After the above rearrangement the matrix $ \tilde{\text{Z}}$ is well-behaved.

%\subsection{Preconditioning and Iterative Solution}
%\label{sec:precond}
%When iterative solvers such as the generalized minimal residual method (GMRES) \cite{saad_gmres_1986}, are used to solve the resulting linear system, they exhibit poor convergence. This is because the singular values of the $\mathcal{T}$ operator comprise two branches, accumulating respectively at zero and at infinity \cite{nedelec_acoustic_2001}. Therefore, over the years various pre-conditioners have been proposed such as incomplete LU factorization \cite{jovanovic_drop_1993,lee_loop_2003,carpentieri_symmetric_2012}, near-zone preconditioners \cite{wiedenmann_effect_2013}, hierarchical basis preconditioners (i.e. multilevel or prewavelet preconditioners) \cite{vipiana_multiresolution_2005,vipiana_multiresolution_2007}. More recently, the  Calderón identity \cite{christiansen_preconditioner_2002} has been used in combination with the Buffa-Christiansen basis functions \cite{buffa_dual_nodate}, thus obtaining a purely  multiplicative preconditioner \cite{andriulli_multiplicative_2008, andriulli_well-conditioned_2013}  so that  its  eigenvalues cluster  at a  certain point far away from the origin and infinity. In this case, we solved the numerical problem by using LU decomposition. The drastic reduction of the number of the unknown and of the dimension of the matrix makes  

\section{Results and Discussion}

The execution of the numerical algorithm can be subdivided into $4$ stages: i) the assembly of the matrices $\text{T}_\pm$, $\text{K}_\pm$ and of the vectors $\text{E}_0$, $\text{H}_0$  in terms of loop/star basis functions; ii) the generation of the longitudinal/transverse static modes of an isolated object by solving the eigenvalue problems \ref{eq:NumericLongProblem} and \ref{eq:NumericTransProblem}; iii) the  ``compression" stage, i.e. passing from the representation of the matrices $\text{T}_\pm$, $\text{K}_\pm$ in terms of the loop/star basis to the representation in terms of the static modes basis. The compression does not need to be performed on the static part of the operators lying on the block-diagonal, since they are diagonalized by the static modes, as shown by Eqs. \ref{eq:Regularization}. iv) Direct matrix inversion using LU decomposition.

\label{sec:Results}

\subsection{Sphere}

\begin{figure}
    \centering
    \includegraphics[width=0.8\columnwidth]{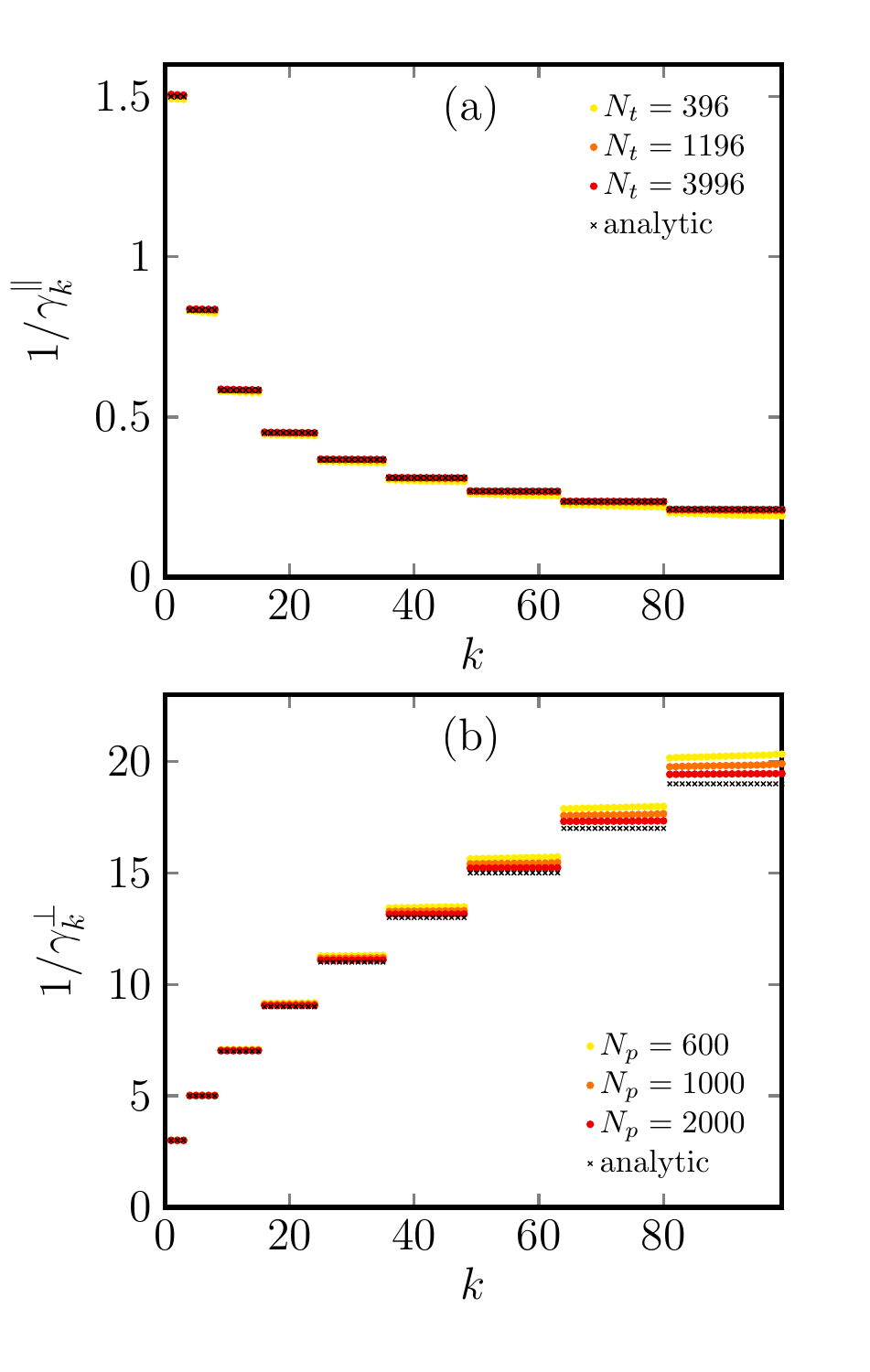}
    \caption{Reciprocal of the first 100 eigenvalues $\gamma_k^\parallel$ (a) and $\gamma_k^\perp$ (b) associated with the longitudinal and transverse static modes of the sphere of unit radius. The eigenvalues are computed for several densities of the triangular surface mesh (filled circles of different colors) having $N_t$ elements and $N_p$ nodes and are compared with their analytical value.}
    \label{fig:Eigenvalues}
\end{figure}

\begin{figure*}
    \centering
    \includegraphics[width=0.9\textwidth]{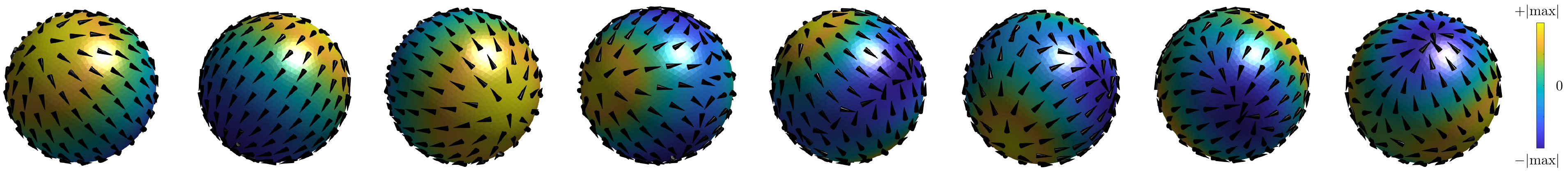}
    \caption{Longitudinal static modes of a sphere. The modes are shown in lexicographic order, sorted (in ascending order)  accordingly to their static eigenvalue. The first 8 modes are shown: the first three are associated to $n=1$ (electric dipole) the next five to $n=2$ (electric quadrupole). The arrows represent the direction of the surface current density field, the colors represent the surface charge density.}
    \label{fig:SphereModeLong}
\end{figure*}

\begin{figure*}
    \centering
    \includegraphics[width=0.9\textwidth]{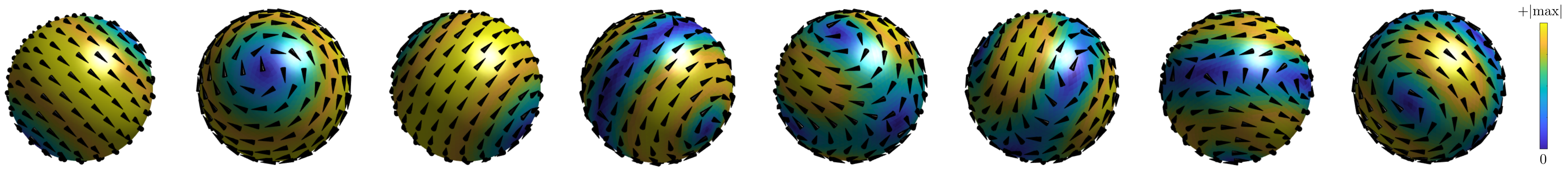}
    \caption{Transverse static modes of a sphere. The modes are shown in lexicographic order, sorted  (in descending order) accordingly to their static eigenvalue. The first 8 modes are shown: the first three are associated to $n=1$ (magnetic dipole) the next five to $n=2$ (magnetic quadrupole). The arrows represent the direction of the surface current density field, the colors represent the magnitude of the current density field.}
    \label{fig:SphereModeTrans}
\end{figure*}

We first consider the scattering from a sphere. This problem has an analytical solution \cite{mie_beitrage_1908,bohren_absorption_1998}.

%\subsubsection{Generation of the static basis}
Figure \ref{fig:Eigenvalues} shows the convergence of the eigenvalues $\{\gamma_k^\parallel \}$ and $\{\gamma_k^\perp \}$ toward their analytical counterpart, given by Eqs. \ref{eq:EigLongSphere} and \ref{eq:EigTransSphere}, as a function of the triangular mesh density. From now on, we consider the static basis calculated using a triangular mesh with $N_p=1000$ nodes and $N_t = 1996$ triangles.  The first 8 longitudinal modes are shown in Fig. \ref{fig:SphereModeLong}, the first 8 transverse modes are shown in Fig. \ref{fig:SphereModeTrans}.

%\subsubsection{Numerical Orthogonality and Gram matrices}

In the discrete problem, the orthogonality between any pair of longitudinal modes and between any pair of transverse modes is always guaranteed, because the matrices $\text{T}^{\circlearrowleft \circlearrowleft}$ and $\text{T}^{\star \star}$ are real and symmetric. This property is indeed verified at the numerical level with machine precision. Instead, even if we expect that the mutual product between a transverse and a longitudinal static mode to be vanishing, this fact is only approximately verified, since the sub-domain basis functions used to represent the longitudinal modes, namely the ``star" functions, are not rigorously curl-free \cite{vecchi_loop-star_1999}. Thus, it is worth calculating the ``mutual" Gram matrix $\text{G}$, whose occurrences are defined as $g_{hk} = \langle \mathbf{j}_h^\perp, \mathbf{\bf j}_k^\parallel \rangle$.
The maximum occurrence of the mutual Gram matrix is 0.024 for a sphere with the considered surface mesh, assuming $\|\mathbf{j}_h^\parallel \|= \|\mathbf{j}_h^\perp\|=1$, $\forall h$.

%In Fig. \ref{fig:Gram100x100} are reported the first $100 \times 100$ occurrences of the Gram Matrix.

\subsubsection{Gold sphere}
We now use the static modes to solve the scattering problem from a gold sphere of radius $R=100$nm in the visible and near infrared spectral range.  We describe the gold permittivity by interpolating experimental data \cite{johnson_optical_1972}. At these frequencies, a metal nano-object may undergo plasmonic resonances, which have an electrostatic origin \cite{mayergoyz_electrostatic_2005}. The sphere is excited by a linearly polarized plane wave of wavelength $\lambda$.

\begin{figure}
    \centering
    \includegraphics[width=\columnwidth]{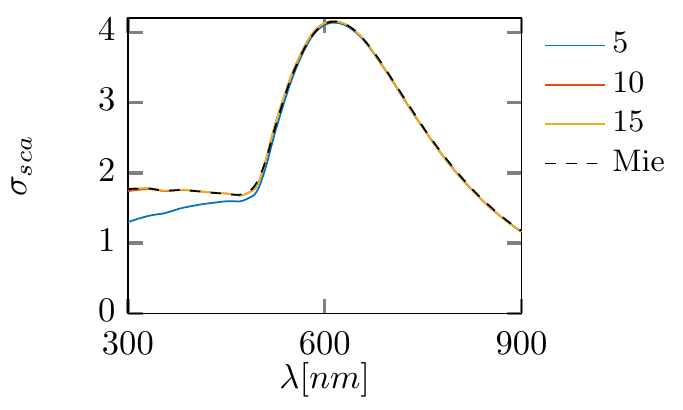}
    \caption{Scattering efficiency $\sigma_{sca}$ of a gold sphere with radius $R=100$nm excited by a linearly polarized plane wave at wavelength $\lambda$. $\sigma_{sca}$ is evaluated using the \texttt{PMCHWT} and using an increasing number of longitudinal and transverse static modes ($\text{N}^\parallel=\text{N}^\perp = 5,10,15$). The reference Mie solution (black dashed line) is also shown for comparison.}
    \label{fig:PlasmonSphere}
\end{figure}

Figure \ref{fig:PlasmonSphere} shows the scattering efficiency $\sigma_{sca}$ as a function of $\lambda$. The scattering efficiency is defined as the scattering cross section normalized by the geometrical cross section $G$ which, in this case, is $G=\pi R^2$ \cite{bohren_absorption_1998}. We consider different solutions, obtained by increasing the number of modes employed in the expansion \ref{eq:SurfaceCurrentExpansion}, by keeping $\text{N}^\parallel=\text{N}^\perp$. The reference Mie solution \cite{bohren_absorption_1998} is also shown for comparison. For $\text{N}^\parallel=\text{N}^\perp = 5$, the numerical solution is in good agreement with the reference solution only in the long-wavelength regime, while it shows a slight disagreement when $\lambda$ becomes comparable with $R$. Increasing the number of modes to $\text{N}^\parallel=\text{N}^\perp = 10$, we obtain a good agreement over the whole investigated spectrum. In this latter case, the inversion of a $40 \times 40$ matrix is required at each frequency.

\begin{figure}
    \centering
    \includegraphics[width=0.7\columnwidth]{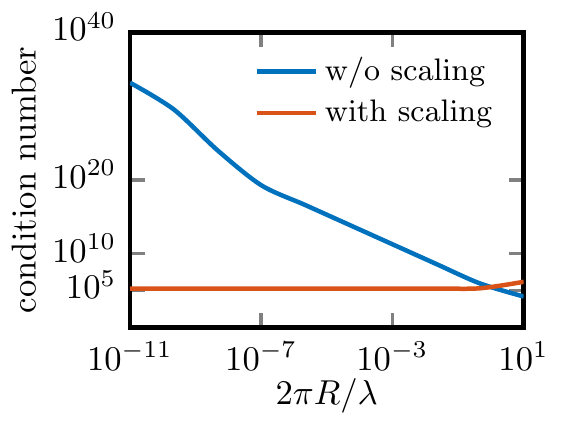}
    \caption{Condition number with and without basis rearrangement and scaling, as a function of the size parameter $2\pi R / \lambda$, assuming $\text{N}^\parallel = \text{N}^\perp=15$.  We considered a gold sphere of  varying radius $R$, excited by a linearly polarized plane wave at wavelength $\lambda=620nm$.}
    \label{fig:ConvergenceGold}
\end{figure}

In Fig. \ref{fig:ConvergenceGold}, we show the condition number of the \texttt{PMCHWT} problem with and without the rearrangement of the basis described in section \ref{sec:LowFreq} as a function of the sphere radius $R$, at $\lambda=620$nm. It is apparent that, without the basis rearrangement,  the condition number exponentially increases, which is symptomatic of the low-frequency breakdown problem. By rearranging the basis, the condition number is constant over the whole investigated range of  $R$.

\begin{figure}
    \centering
    \includegraphics[width=0.8\columnwidth]{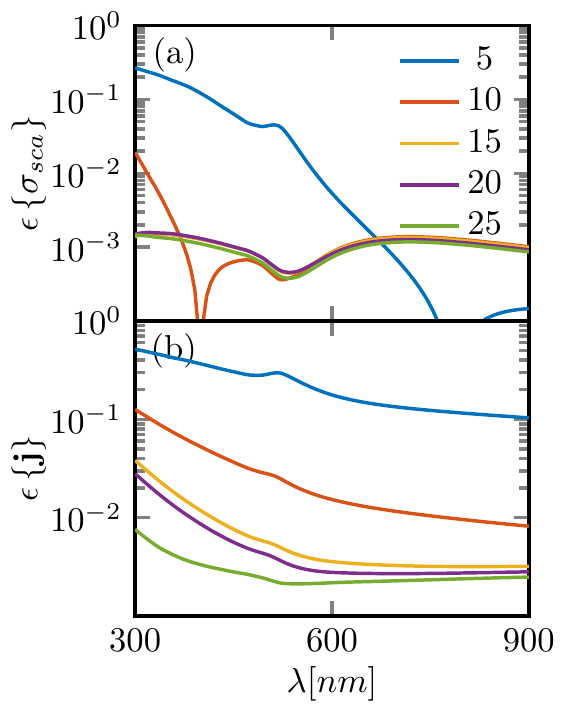}
    \caption{Error made in the evaluation of the scattering efficiency $\epsilon[\sigma_{sca}]$ (a) and of the equivalent surface currents $\epsilon[\mathbf{j}]$ (b) of a gold sphere $R=100$nm by using $\text{N}^\parallel=\text{N}^\perp = 5,10,15,25$ static modes ($20,40,100$ degrees of freedom). The reference loop/star solutions are obtained using $999$ loop and $1996$ star functions ($5998$ degrees of freedom). In both cases a \texttt{PMCHWT} formulation is used. The sphere is excited by a linearly polarized plane wave as a function of the wavelength $\lambda$.}
    \label{fig:Error_Plasmon}
\end{figure}

We now present a more systematic error analysis. In particular, we define the relative error on the scattering efficiency as
\begin{equation}
    \epsilon \left[ \sigma_{sca} \right] = {\left| \sigma_{sca} - \tilde{\sigma}_{sca} \right|}\, / \, {\tilde{\sigma}_{sca}},
\end{equation}
where $\tilde{\sigma}_{sca}$ is the reference solution which is obtained by solving the \texttt{PMCHWT} problem applying the finite element method with $N^\circlearrowleft=999$ loop functions and $N^\star=1995$ star functions, associated with the same mesh used for the static modes generation. In Fig. \ref{fig:Error_Plasmon} (a), we plot $\epsilon \left[ \sigma_{sca} \right]$ as a function of the wavelength $\lambda$, by varying the number of static modes, keeping $\text{N}^\parallel=\text{N}^\perp$. We note that for $\text{N}^\parallel=\text{N}^\perp \ge 10$ the achieved error is lower than $0.002$ all over the investigated spectral range. The error only slowly decrease if the number of modes $\text{N}^\parallel=\text{N}^\perp$ is increased from $15$ to $25$. Then, we investigate the error in the evaluation of the equivalent surface currents. They are immediately related to the total electric field on the surface of the object, which has a great importance in nano-optics applications \cite{schuller_plasmonics_2010}. We define the relative error as:
\begin{equation}
      \epsilon \left[ \text{J} \right] = {\left\| \text{J} - \tilde{\text{J}} \right\|_2}/{\left\|\tilde{\text{J}}_e\right\|_2},
\end{equation}
where $\tilde{\text{J}}$ is the reference loop/star solutions and $\left\| \cdot \right\|_2$ is the Euclidean norm. In Fig. \ref{fig:Error_Plasmon} (b), we show $\epsilon \left[ \text{J} \right]$ as a function of $\lambda$. Compared to panel (a), it is apparent that more static basis functions are needed to achieve a prescribed error.

\begin{figure}
    \centering
    \includegraphics[width=\columnwidth]{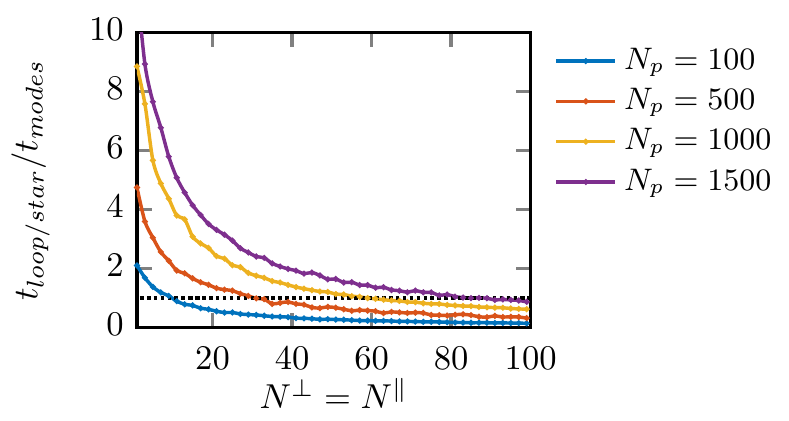}
    \caption{Ratio between the total cpu time for the computation of the \texttt{PMCHWT} solution using the loop/star basis $t_{loop/star}$ and using the static mode basis $t_{modes}$. This ratio is evaluated  as a function of the number of static modes $\text{N}^\perp=\text{N}^\parallel$ for different meshes with different number of nodes $N_p$, for the problem of scattering by a gold sphere excited by a plane wave with $\lambda=620nm$.} 
    \label{fig:SpeedUP}
\end{figure}

We now analyze the cpu time required for the different stages of the numerical solution of the \texttt{PMCHWT} using static modes. The time spent by the algorithm for the  matrix inversion using LU decomposition and for the static mode calculation is always negligible with respect to the time required for the assembly of the matrices $\text{K}_\pm$ and $\text{T}_\pm$ and for their compression. The compression is typically the most time consuming stage, and its cpu time increases linearly with the number of transverse/longitudinal modes employed.
In Fig. \ref{fig:SpeedUP}, we show the ratio of the cpu-time   $t_\mathtt{loop/star}$ required to obtain the solution of the \texttt{PMCHWT} using the loop/star basis  to the cpu-time $t_\mathtt{modes}$  required to obtain the solution using  $\text{N}^\perp=\text{N}^\parallel$ static modes. This analysis is repeated for several mesh densities. The same mesh is used both for the calculation of the static modes and for the loop/star solution.  For a surface mesh with $N_p=100$ nodes, $t_\mathtt{loop/star}>t_\mathtt{modes}$ only when a few modes are used, namely $\text{N}^\perp=\text{N}^\parallel < 10$; for a number of modes larger than this threshold, the loop/star solution becomes faster. For denser meshes, the use of the static basis may become more favorable, while the condition $t_\mathtt{loop/star}=t_\mathtt{modes}$ is verified for a larger number of basis functions employed. For the mesh density used in the previous examples with $N_p=1000$, the static mode solution is faster for $\text{N}^\perp=\text{N}^\parallel < 69$.

\begin{figure}
    \centering
    \includegraphics[width=\columnwidth]{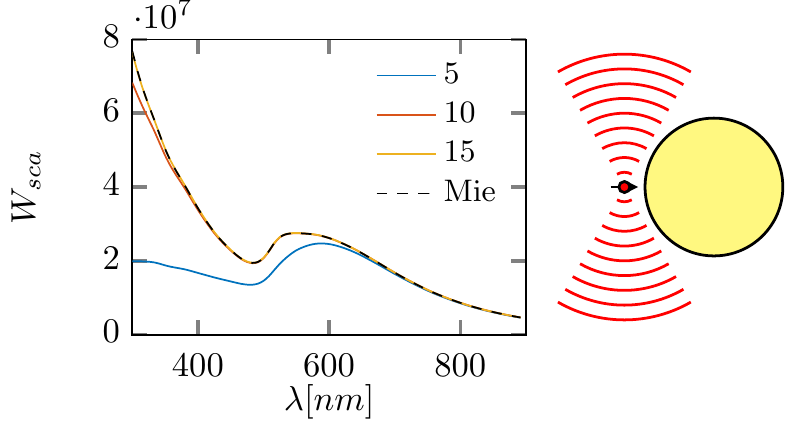}
    \caption{Scattered power $W_{sca}$ by a gold sphere with radius $R=100$nm excited by an electric point dipole, located at a distance of $30$nm from the surface of the particle, and oriented as shown in the inset. $W_{sca}$ is evaluated as a function of the wavelength $\lambda$ using the \texttt{PMCHWT} and employing an increasing number of longitudinal/transverse static modes ($\text{N}^\parallel=\text{N}^\perp = 5,10,15$). The reference Mie solution (black dashed line) is also shown for comparison.}
    \label{fig:PlasmonSphereDipole}
\end{figure}

We now investigate the convergence of the static modes solution in the presence of an excitation located close to the object's surface. In Fig. \ref{fig:PlasmonSphereDipole}, we consider an electric point dipole exciting a gold sphere of radius $R=100$nm. The point dipole is placed at a distance of $30$nm from the sphere's surface and oriented as sketched in the inset. We evaluate the total scattered power $W_{sca}$ using an increasing number of longitudinal/transverse static modes ($\text{N}^\parallel=\text{N}^\perp = 5,10,15$). We also show the reference Mie solution (black dashed line) for comparison. As soon as $\text{N}^\parallel=\text{N}^\perp = 15$, the two solutions become indistinguishable. 

\begin{figure}
    \centering
    \includegraphics[width=1\columnwidth]{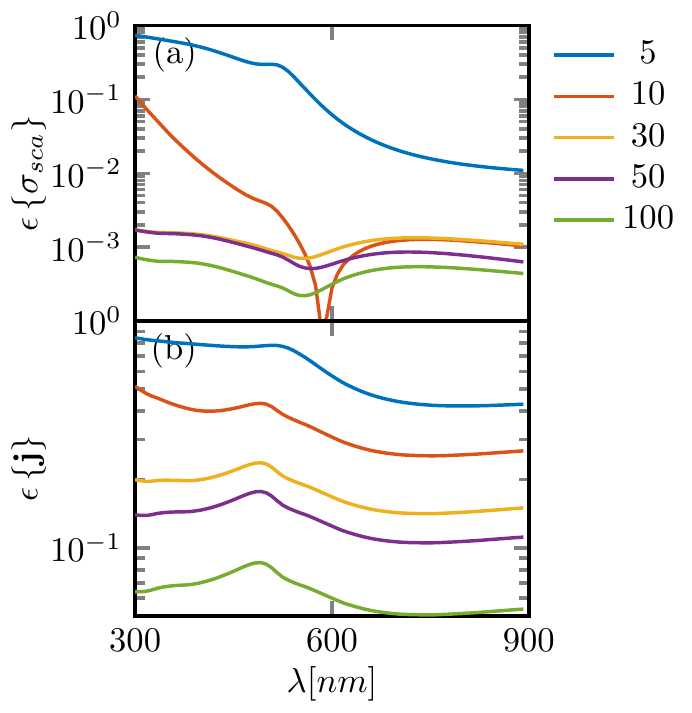}
    \caption{Error made in the evaluation of the scattered power $W_{sca}$ and of the equivalent surface currents $\epsilon[\mathbf{j}]$ (b)  of a gold sphere $R=100$nm by solving the \texttt{PMCHWT} using $\text{N}^\parallel=\text{N}^\perp = 5,10,30,50,100$ static modes. The reference quantities are obtained by solving the \texttt{PMWCHWT} using  $999$ loop and $1996$ star functions. The sphere is excited by a dipole, located at a distance of $30$nm from the surface of the particle and oriented as shown in the inset of Fig. \ref{fig:PlasmonSphereDipole}.}
    \label{fig:Error_Plasmon_Dipole}
\end{figure}

In Fig. \ref{fig:Error_Plasmon_Dipole}, we investigate the two relative errors $\epsilon[W_{sca}]$ and $\epsilon[\mathbf{j}]$, assuming the loop/star solution  as reference.  Figure \ref{fig:Error_Plasmon_Dipole} (a) shows that the total scattered power evaluated using the static modes quickly converges toward the loop/star solution, with a rate comparable to the one observed in Fig. \ref{fig:Error_Plasmon} (a). Instead, the convergence of the surface currents is much slower because rapid spatial variations of the surface current in proximity of the exciting dipole require higher order static modes to be accurately described.

\subsubsection{High-permittivity sphere} 

\begin{figure}
    \centering
    \includegraphics[width=\columnwidth]{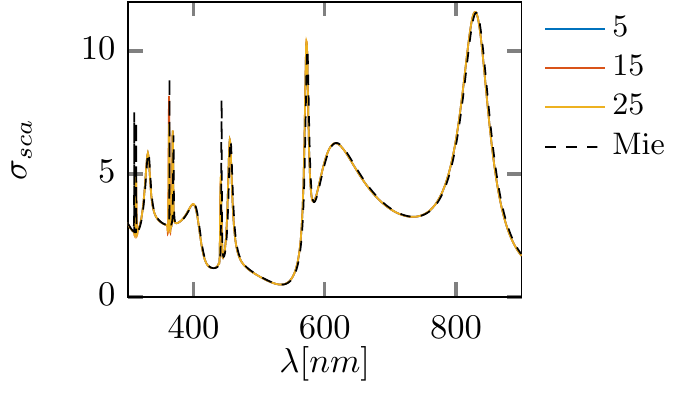}
    \caption{Scattering efficiency $\sigma_{sca}$ of a dielectric sphere with $R=100$nm and $\varepsilon_R = 16$, excited by a linearly polarized plane wave evaluated by the \texttt{PMCHWT} with an increasing number of longitudinal and transverse static modes ($\text{N}^\parallel=\text{N}^\perp = 5,10,15$). The reference Mie solution (black dashed line) is also shown for comparison.}
    \label{fig:CscaSphereDiel}
\end{figure}

We now consider a high permittivity sphere, assumed to be non-dispersive in time with relative permittivity $\varepsilon_R=16$. Sub-wavelength objects of sufficiently high permittivity may support scattering resonances, which have a magnetostatic origin \cite{forestiere_magnetoquasistatic_2020,forestiere_resonance_2020}.

In Fig. \ref{fig:CscaSphereDiel}, we calculate the scattering efficiency $\sigma_{sca}$ as a function of the wavelength of the exciting, linearly polarized, plane wave. We consider different solution computed using an increasing the number of modes, by keeping $\text{N}^\parallel=\text{N}^\perp$. We use as reference the analytic Mie solution \cite{mie_beitrage_1908,bohren_absorption_1998}, shown with a black dashed line. Even if a good agreement is found at low frequency with $\text{N}^\parallel=\text{N}^\perp=5$, the accuracy of the solution  deteriorates at higher frequencies, and it is unable to describe some peaks of the scattering response. By increasing the number of employed modes to $\text{N}^\parallel=\text{N}^\perp=10$, we obtain a very good agreement over the whole investigated spectrum.

\begin{figure}
    \centering
    \includegraphics[width=0.9\columnwidth]{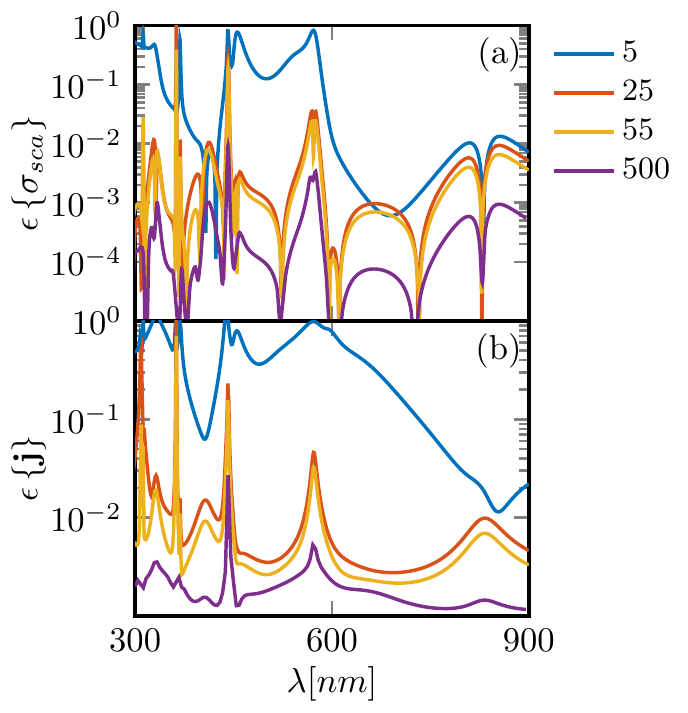}
    \caption{Error made in the evaluation of the scattering efficiency $\sigma_{sca}$ (a) and of the equivalent surface currents $\mathbf{j}$ (b) of a dielectric sphere with $\varepsilon_R=16$ and $R=100$nm by solving the \texttt{PMCHWT} equation using the static mode basis with $\text{N}^\parallel=\text{N}^\perp = 5,25,55,500$. The reference values are obtained by solving the  \texttt{PMCHWT}  and using the loop/star expansion. The sphere is excited by a linearly polarized plane wave as a function of the wavelength $\lambda$.}
    \label{fig:ErrorSphereDiel}
\end{figure}

We now quantify the errors in the scattering efficiency and in the surface currents of the static mode solution. In Fig. \ref{fig:ErrorSphereDiel} (a), we plot $\epsilon \left[ \sigma_{sca} \right]$ as a function of  $\lambda$, employing an increasing number of modes with the constraint $\text{N}^\perp=\text{N}^\parallel$. When $\text{N}^\perp=\text{N}^\parallel=5$, the error is acceptable as long $\lambda$ is much larger than the dimension of the object, then the error suddenly increases for wavelengths below $600$nm. By assuming $\text{N}^\perp=\text{N}^\parallel=25$, we obtain a low error all over the investigated spectrum. Only at the resonances the error slightly exceeds $0.01$. A further increase in the number of basis function improves the convergence, especially in the neighborhood of the resonance peaks. In order to check the convergence of the static mode expansion, we also considered $\text{N}^\perp=\text{N}^\parallel=500$. We obtain errors which are less than $0.009$ for the $\sigma_{sca}$ and less than $0.03$ for $\mathbf{j}$ over the whole investigated spectral range.

\subsection{Rod}
\begin{figure*}
    \centering
    \includegraphics[width=\textwidth]{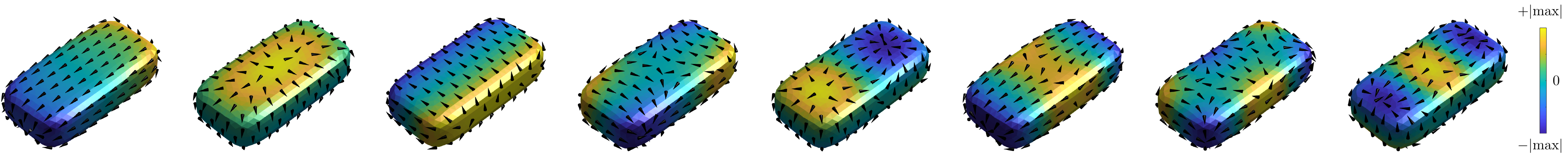}
    \caption{Longitudinal static modes of a rod with semi-axis $1:0.5:0.25$. The modes are shown in lexicographic order (ascending), sorted accordingly to their static eigenvalue. The first 8 modes are shown. The arrows represent the direction of the surface current density field, the colors represent the surface charge density.}
    \label{fig:RodModeLong}
\end{figure*}

\begin{figure*}
    \centering
    \includegraphics[width=\textwidth]{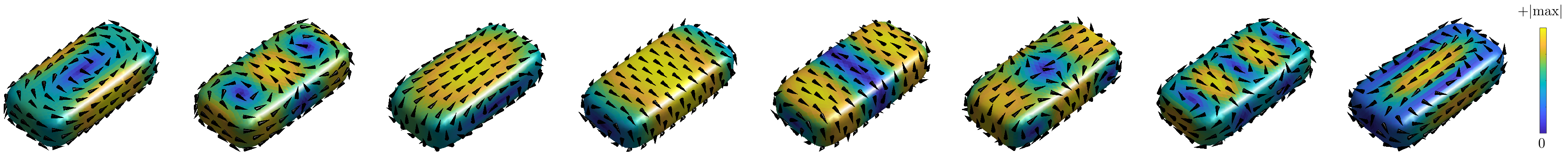}
    \caption{Transverse static modes of a rod with semi-axis $1:0.5:0.25$. The modes are shown in a lexicographic order (descending), sorted accordingly to their static eigenvalue. The first 8 modes are shown. The arrows represent the direction of the surface current density field, the colors represent the magnitude of the current density field.}
    \label{fig:RodModeTrans}
\end{figure*}

We now consider a non-canonical shape, namely a three-dimensional rod. We model this shape as a superellipsoid, whose boundary has the implicit equation
$ \left({x}/{a} \right)^r +  \left({y}/{b} \right)^r +  \left({z}/{a} \right)^r = 1, $
with $b=0.5a$, $c=0.25a$, and $r=6$. We used the public domain code developed by Per-Olof Persson and Gilbert Strang \cite{persson_simple_2004} to generate a surface mesh with $1000$ nodes and $1996$ triangular elements. The first 8 longitudinal and 8 transverse static modes are shown in Fig. \ref{fig:RodModeLong} and \ref{fig:RodModeTrans}, respectively. 

\subsubsection{Gold rod}

\begin{figure}
    \centering
    \includegraphics[width=\columnwidth]{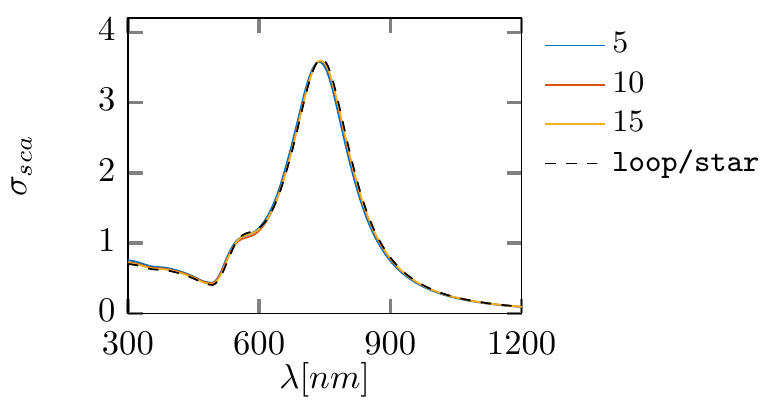}
    \caption{Scattering efficiency $\sigma_{sca}$ of a gold rod with semi-axis $a=100nm$, $b=0.5a$, and $c=0.25a$ evaluated with an increasing number of longitudinal and transverse static modes $\text{N}^\parallel=\text{N}^\perp = 5,15,25,35,55$. The rod is excited by a linearly polarized plane wave.  The reference loop/star solution (black dashed line) is also shown for comparison.}
    \label{fig:CscaPlasmonRod}
\end{figure}

\begin{figure}
    \centering
    \includegraphics[width=0.9\columnwidth]{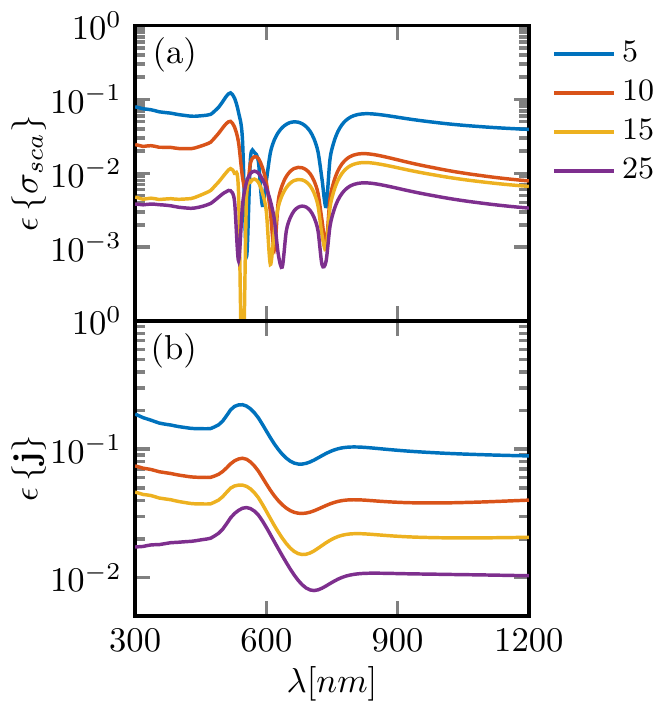}
    \caption{Error $\epsilon$ in the evaluation of the scattering efficiency $\sigma_{sca}$ and of the equivalent surface current density of a gold rod with semi-axis $a=100nm$, $b=0.5a$, and $c=0.25a$ by solving the \texttt{PMCHWT} using the static mode basis with $\text{N}^\parallel=\text{N}^\perp = 5,15,20,25$.  The reference values are obtained by solving the  \texttt{PMCHWT}  using the loop/star expansion. }
    \label{fig:ErrorPlasmonRod}
\end{figure}

First, we investigate a gold \cite{johnson_optical_1972} rod with $a=100$nm, excited by a plane wave linearly polarized along the direction $ \left({\hat{\bf x} + \hat{\bf y}}\right)/{\sqrt{2}}$ and propagating along the $\hat{\bf z}$ axis.  In Fig. \ref{fig:CscaPlasmonRod} we plot the spectrum of the scattering efficiency $\sigma_{sca}$, obtained by increasing the number of modes $\text{N}^\perp=\text{N}^\parallel=5,10,15$. We take as reference the loop/star solution, with $5984$ total degrees of freedom. Only five longitudinal and transverse modes ($20$ total degrees of freedom) are sufficient to achieve a sufficiently good agreement with the reference solution over the whole investigated spectrum, demonstrating a drastic reduction of the total number of unknowns.

In Fig. \ref{fig:ErrorPlasmonRod}, we perform a more systematic analysis of the error on the scattering cross section $\epsilon \left[ \sigma_{sca} \right]$ (a), and on the equivalent surface currents $\epsilon \left[ \mathbf{j} \right]$ (b). Overall, the errors are slightly higher than in the case of a sphere. Besides that, as in the previous numerical experiments, we conclude that i) the error on the surface currents are one order of magnitude higher then the error on the scattering efficiency, and ii) the rate of convergence is not uniform as a function of the number of employed modes, and becomes slower as $\text{N}^\parallel=\text{N}^\perp$ increases.

\subsubsection{High-permittivity Rod}
\begin{figure}
    \centering
    \includegraphics[width=\columnwidth]{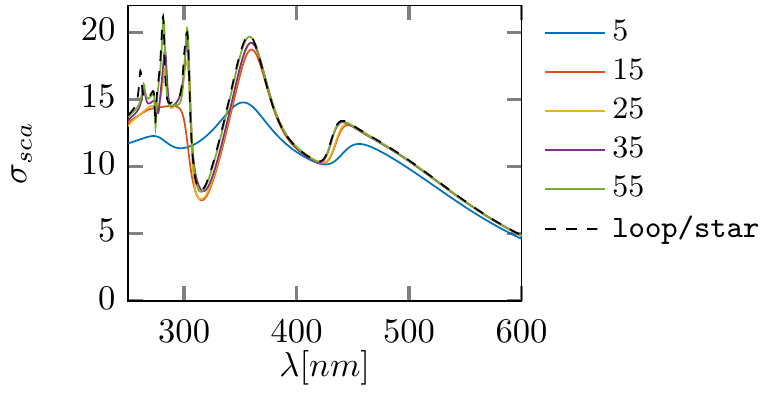}
    \caption{Scattering efficiency $\sigma_{sca}$ of a dielectric rod with $\varepsilon_R=16$ and semi-axis $a=100nm$, $b=0.5a$, and $c=0.25a$ evaluated with an increasing number of longitudinal and transverse static modes $\text{N}^\parallel=\text{N}^\perp = 5,15,25,35,55$. The rod is excited by a linearly polarized plane wave.  The reference loop/star solution (black dashed line) is also shown for comparison.}
    \label{fig:CscaDielRod}
\end{figure}

\begin{figure}
    \centering
    \includegraphics[width=0.9\columnwidth]{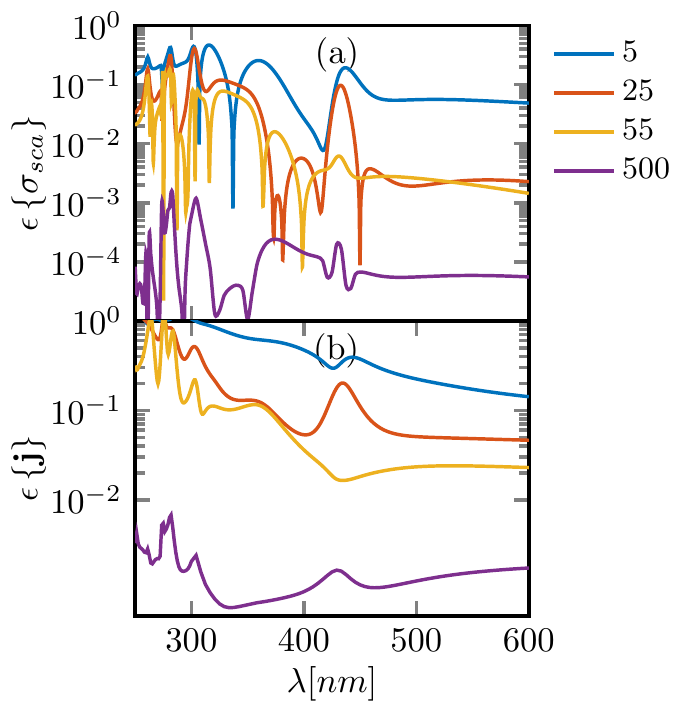}
    \caption{Error $\epsilon$ in the evaluation of the scattering efficiency $\sigma_{sca}$ and of the equivalent surface current density of a dielectric rod with $\varepsilon_R=16$ with semi-axis $a=100nm$, $b=0.5a$, and $c=0.25a$ by solving the \texttt{PMCHWT} using an increasing number of longitudinal and transverse static modes $\text{N}^\parallel=\text{N}^\perp = 5,25,55,500$. The reference values are obtained by solving the  \texttt{PMCHWT}  using the loop/star expansion.}
    \label{fig:ErrorDielRod}
\end{figure}

Here, we investigate a high-permittivity rod. The relative permittivity of the rod is assumed constant over the investigated frequency spectrum to the value $\varepsilon_R = 16$. In Fig. \ref{fig:CscaDielRod}, we plot the $\sigma_{sca}$ spectrum, obtained by increasing the numbers of static modes employed. We use as reference the loop/star solution, with $5984$ degrees of freedom. If the wavelength is much larger than the dimension of the rod,  $\text{N}^\perp=\text{N}^\parallel=15$ are enough to correctly describe the scattering cross section. Nevertheless, the accuracy is lost as soon as the wavelength becomes comparable to the linear dimensions of the rod and high-frequency resonance peaks are not correctly described. Only by increasing the number of modes $\text{N}^\perp=\text{N}^\parallel$ to $55$, all the resonance peaks, including the high-frequency ones, are correctly described.

Figure \ref{fig:ErrorDielRod} offers a more quantitative analysis of the errors  (a) $\epsilon \left[ \sigma_{sca} \right]$, (b) $\epsilon \left[ \mathbf{j} \right]$. We note that, while for large wavelength a number of modes $\text{N}^\perp=\text{N}^\parallel=15$ is sufficient to have an error below $\epsilon \left[ \sigma_{sca} \right]<0.01$ and $\epsilon \left[ \mathbf{j} \right]<0.2$, if the wavelength becomes comparable to the rod largest dimension, as many as $55$ longitudinal and transverse static modes are needed to contain the error. Even in this case, the accuracy is deteriorated in the neighborhood of the resonance peaks. In order to check the convergence of the static mode expansion, we also considered $\text{N}^\perp=\text{N}^\parallel=500$. We obtain errors which are less than $0.0015$ for the $\sigma_{sca}$ and less than $0.0068$ for $\mathbf{j}$ over the whole investigated spectral range.

In conclusion, even if compared to the previously investigated scenarios, a higher number of modes is needed to correctly describe the unknown current densities, the number of modes needed to describe the solution still remains much smaller than the number of loop/star functions required to achieve a comparable accuracy.

\subsection{Solution of Multiple Scattering Problems}

\begin{figure}
    \centering
    \includegraphics[width=\columnwidth]{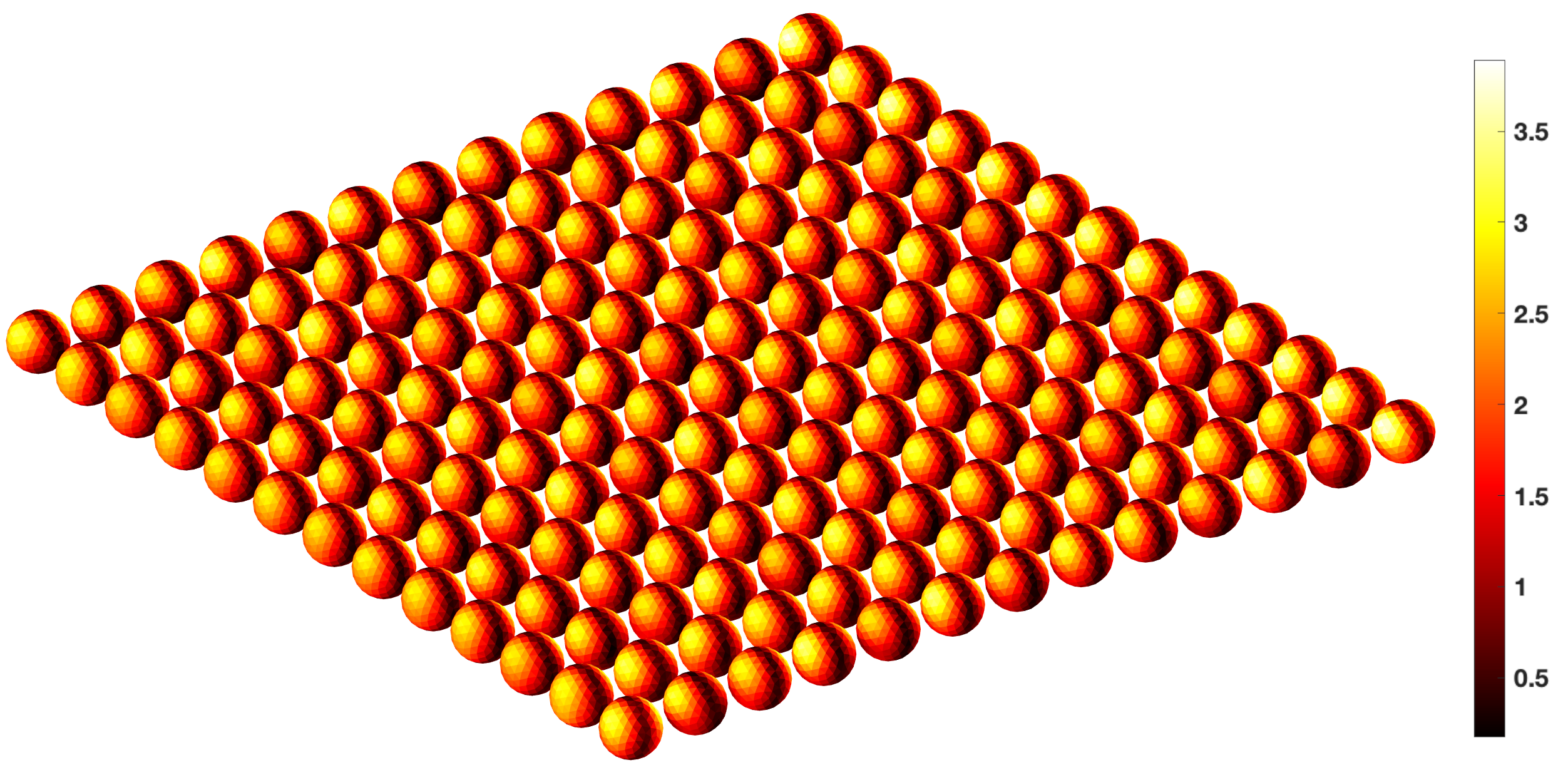}
    \caption{Magnitude of the electric field on the particles' surface of a finite-size $13\times 13$ periodic array of $169$ gold spheres with radius $R=100nm$ and edge-edge distance $50nm$. The edge of the array is $3.2 \mu m$. The array is excited by a linearly polarized plane of wavelength $\lambda=600nm$.  The solution has been computed by using the static mode expansion with $\text{N}^\parallel=\text{N}^\perp=10$.}
    \label{fig:peridic_array_E}
\end{figure}

\begin{figure}
    \centering
    \includegraphics[width=0.9\columnwidth]{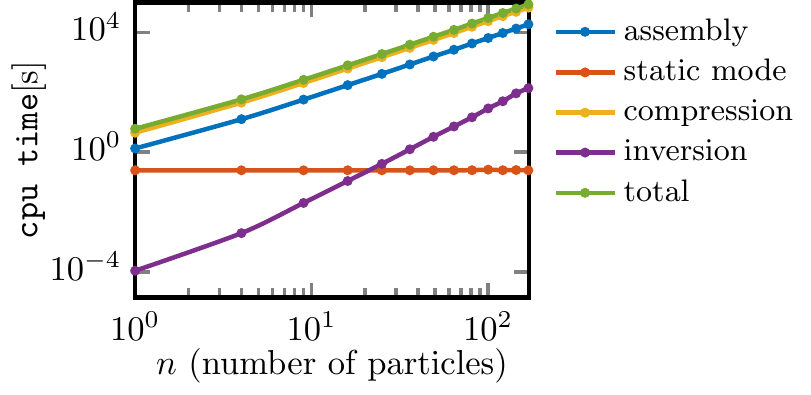}
    \caption{CPU time (in seconds) of the different stages of the numerical solution of the scattering problem from an array of $n$ spheres:  Assembly of the matrices using the loop/star basis. Generation of the longitudinal/transverse static modes. Compression, by passing from loop/star to static modes. Direct inversion (LU). Total Time. The code is implemented in \texttt{FORTRAN}, and run on a single cpu.}
    \label{fig:TimeArray}
\end{figure}

\begin{figure}
    \centering
    \includegraphics[width=\columnwidth]{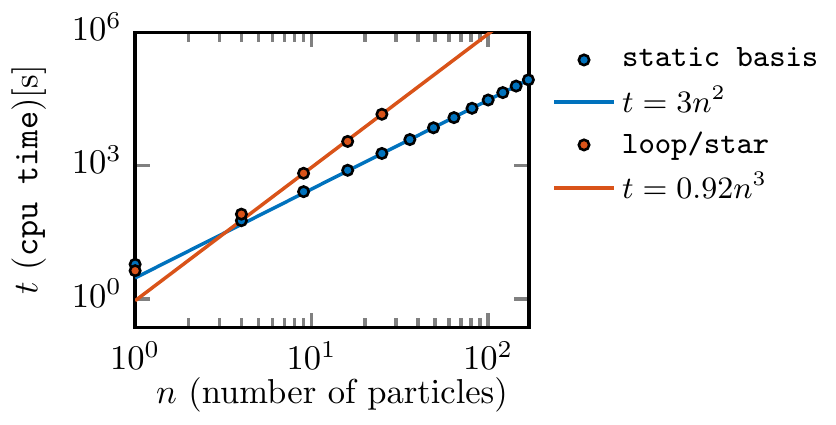}
    \caption{Total execution CPU time (in seconds) of the numerical solution of the scattering problem from an array of $n$ spheres by using loop/star basis functions and by usign the static modes basis. Both codes are implemented in \texttt{FORTRAN} and run on a single cpu. The two fitting curves are shown with a continuous line.}
    \label{fig:TimeComparison}
\end{figure}

The static mode basis finds its natural application in the numerical solution of multiple scattering problems, where an array, whose dimension can be much larger than the incident wavelength, is made of $n$ objects with identical shape but with different orientation and size. 
\begin{table}[]
    \centering
    \begin{tabular}{c|cccccc}
    $n$ & 1 & 4 & 9 & 16  & 25 \\
    $\epsilon[\sigma_{sca}] [\%]$ & 0.2 & 0.68 & 2.5 & 2.8 & 3.3 \\
    $\epsilon[\mathbf{j}_{sca}]$ & 2.0 & 3.3 & 3.4 & 3.5 & 3.4
    \end{tabular}
    \caption{Errors in the solution of the scattering problem by a periodic array made by $n$ spheres.}
    \label{tab:my_label}
\end{table}

We support the above statement through several examples. Specifically, we consider a finite-size periodic array of $n$ spheres, of radius $R=100nm$, which are placed at the nodes of a $\sqrt{n}\times \sqrt{n}$ square grid of pitch $250$nm. The values of pitch and radius also fix the edge-edge interparticle separation to $50$ nm. The geometry of a $13 \times 13$ array of $169$ spheres is shown in Fig. \ref{fig:peridic_array_E}. The array is excited by a plane wave, linearly polarized in the plane of the array, along one of the two axis, and propagating in the direction orthogonal to the array's plane with wavelength $\lambda=600$nm. Each sphere is described by a surface mesh having $200$ nodes and $396$ triangles and $594$ edges, which corresponds to a maximum edge length of the mesh's triangles $\approx\lambda/17$. A loop/star description leads to $1188 \, n$ unknowns. Instead, we considered $\text{N}^\parallel$ longitudinal and $\text{N}^\perp=10$ transverse modes to describe the electric and magnetic surface currents on each nanoparticle, which correspond to $40 \, n $ unknowns ( a reduction of more than one order of magnitude). 

In Fig. \ref{fig:TimeArray}, we show the cpu time (in seconds) of the different stages of the numerical solution of the scattering problem from the finite-size periodic array of spheres, as a function of the number $n$ of spheres ($n=1,4,9,16,\ldots,169$). The \texttt{FORTRAN} code runs on a single cpu. %The time required by each stage during our numerical experiment is shown together with the total computational time as a function of the number of particles in the array.
The total computational time is always dominated by the compression time, which scales as $\propto n^2$. The assembly time, which also scales as $\propto n^2$, gives an important contribution to the total execution time. The time required for the computation of the static modes of the isolated sphere, which are used as basis, is negligible and  it does not depend on the array's size. The inversion time, which scales as $\propto n^3$, is negligible if compared with the compression time, as in our case, if the dimensions of the arrays do not exceed a given number of particles ($n_{th}$). The value of $n_{th}$ can be estimated extrapolating the the curves in Fig. \ref{fig:TimeArray}, and it is $n_{th}\approx 80k$ particles.

In Fig. \ref{fig:TimeComparison} we compare the total cpu-time required for the solution of the \texttt{PMCHWT} solution using the static modes basis against the corresponding time required to obtain the loop/star solution, as a function of $n$. For $n>4$, the static mode solution becomes advantageous, and the speed up further increases as $n$ increases. In Tab. \ref{tab:my_label}, we also compare for $n\le25$ the error made by the static mode solution compared to the loop/star reference: it is always less then $4 \%$ for both the scattering efficiency and the currents.

To obtain information on the scaling laws, we fit the total computational time of the solution of the \texttt{PMCHWT} in terms of the static basis using the curve $t = 3n^2$, and in terms of the loop/star solution using the curve $t=0.92n^3$. The ratio is $0.3 n $. Thus, for a $5\times 5$ array, the cpu-time required to obtain the loop/star solution is $7.5$ time larger than the cpu-time required for the static mode solution. For $n=169$ we extrapolate that the static mode solution is $50 \times$ faster, but it was not possible to verify it experimentally, since the time expected to compute the loop/star solution was prohibitive.

In Fig. \ref{fig:peridic_array_E}, we plot the magnitude of the total electric field on the surface of each sphere of the $n=169$ sphere array. For this case  $n=169$, we consider $520$ static modes. The corresponding loop/star description would require $200k$ unknowns.

\section{Conclusions} 
\label{sec:Conclusion}
We introduced a set of ``static" surface current modes and used them to expand the unknown surface current densities in the surface integral equations governing the electromagnetic scattering problem from penetrable objects. We demonstrated the effectiveness of the static mode expansion in the solution of the Poggio-Miller-Chang-Harrington-Wu-Tsai (\texttt{PMCHWT}) formulation  \cite{wu_scattering_1977,chang_surface_1977,poggio_chapter_1973}.
We found several characteristics that make the use of static modes appealing:
\begin{itemize}
    \item The retarded Green's function, constituting the kernel of integral operators recurring in surface integral formulations such as the \texttt{PMCHWT}, may be decomposed as the sum of the static Green's function (with integrable singularity) and a proper difference (which is a regular function); the resulting integral operators containing the static Green's function are diagonalized by the static modes, thus the overall problem is regularized \cite{nosich_method_1999}.
    \item The use of the static mode expansion combined with an appropriate rescaling and rearranging of the unknowns makes the \texttt{PMCHWT} formulation immune from the low-frequency breakdown problem.
    \item The static modes only depend on the shape of the object, thus, the same static basis can be used (and has the same advantages) regardless of the operating frequency and material of object. This fact enables the description of any scattering scenario involving one of more  objects of a given shape, in terms of the same  ``alphabet'' of basis functions, regardless of the frequency of operation. This fact constitutes an advantage compared to other basis sets (e.g., characteristic modes), where the modes depend on frequency and materials, thus the {alphabet of basis functions}, in which the scattering process is described, changes every time one of these parameters is varied, thus preventing a unified description.
\end{itemize}

 As test cases, we considered the scattering from both metal and high-permittivity dielectric objects. We found that for objects of size smaller than the wavelength of operation, excited by a slowing varying electromagnetic field, only few modes are sufficient to accurately describe the emergent scattering response. Thus, the use of the static mode expansion drastically reduces the number of unknowns compared to a discretization in terms of loop/star or RWG functions, without deteriorating the accuracy of the solution.

The most time consuming stage of the \texttt{PMCHWT} numerical solution using static modes is typically the ``compression" stage, which is the change of the basis used to represent the impedance matrix from the loop/star set to the static modes set.  In the scattering problem from an isolated particle,  the use of the static modes set is convenient in terms of total cpu-time, compared with the use of a loop/star set, only when few static modes are needed to correctly describe the surface currents. Instead, the use of the static modes is very convenient in multiple scattering  problems from particle's arrays made of $n$ objects with identical shape but different orientation and size. Even for small arrays the solution in terms of static mode is faster then the solution in terms of loop/star, and this advantage linearly increases with the number of particles $n$.

In conclusions, the results presented in this paper may promote the use of the static basis in multiple scattering problems, including the numerical modeling of metasurfaces and metalens. 

% Generated by IEEEtran.bst, version: 1.14 (2015/08/26)


\begin{thebibliography}{10}
\providecommand{\url}[1]{#1}
\csname url@samestyle\endcsname
\providecommand{\newblock}{\relax}
\providecommand{\bibinfo}[2]{#2}
\providecommand{\BIBentrySTDinterwordspacing}{\spaceskip=0pt\relax}
\providecommand{\BIBentryALTinterwordstretchfactor}{4}
\providecommand{\BIBentryALTinterwordspacing}{\spaceskip=\fontdimen2\font plus
\BIBentryALTinterwordstretchfactor\fontdimen3\font minus
  \fontdimen4\font\relax}
\providecommand{\BIBforeignlanguage}[2]{{%
\expandafter\ifx\csname l@#1\endcsname\relax
\typeout{** WARNING: IEEEtran.bst: No hyphenation pattern has been}%
\typeout{** loaded for the language `#1'. Using the pattern for}%
\typeout{** the default language instead.}%
\else
\language=\csname l@#1\endcsname
\fi
#2}}
\providecommand{\BIBdecl}{\relax}
\BIBdecl

\bibitem{collin_antennas_1985}
\BIBentryALTinterwordspacing
R.~E. Collin, \emph{Antennas and {Radiowave} {Propagation}}.\hskip 1em plus
  0.5em minus 0.4em\relax McGraw-Hill College, 1985. [Online]. Available:
  \url{https://books.google.it/books/about/Antennas_and_Radiowave_Propagation.html?id=YHepAAAACAAJ&redir_esc=y}
\BIBentrySTDinterwordspacing

\bibitem{yu_flat_2014}
\BIBentryALTinterwordspacing
N.~Yu and F.~Capasso, ``\BIBforeignlanguage{en}{Flat optics with designer
  metasurfaces},'' \emph{\BIBforeignlanguage{en}{Nature Materials}}, vol.~13,
  no.~2, pp. 139--150, Feb. 2014, bandiera\_abtest: a Cg\_type: Nature Research
  Journals Number: 2 Primary\_atype: Reviews Publisher: Nature Publishing Group
  Subject\_term: Metamaterials Subject\_term\_id: metamaterials. [Online].
  Available: \url{https://www.nature.com/articles/nmat3839}
\BIBentrySTDinterwordspacing

\bibitem{khorasaninejad_achromatic_2015}
\BIBentryALTinterwordspacing
M.~Khorasaninejad, F.~Aieta, P.~Kanhaiya, M.~A. Kats, P.~Genevet, D.~Rousso,
  and F.~Capasso, ``Achromatic {Metasurface} {Lens} at {Telecommunication}
  {Wavelengths},'' \emph{Nano Letters}, vol.~15, no.~8, pp. 5358--5362, Aug.
  2015. [Online]. Available: \url{https://doi.org/10.1021/acs.nanolett.5b01727}
\BIBentrySTDinterwordspacing

\bibitem{chew_fast_2000}
W.~C. Chew, J.-M. Jin, E.~Michielssen, and J.~Song, Eds.,
  \emph{\BIBforeignlanguage{English}{Fast and {Efficient} {Algorithms} in
  {Computational} {Electromagnetics}}}.\hskip 1em plus 0.5em minus 0.4em\relax
  Boston: Artech House, Jul. 2000.

\bibitem{bucci_use_1995}
\BIBentryALTinterwordspacing
O.~M. Bucci and G.~D. Massa, ``Use of characteristic modes in
  multiple-scattering problems,'' \emph{Journal of Physics D: Applied Physics},
  vol.~28, no.~11, p. 2235, Nov. 1995. [Online]. Available:
  \url{https://iopscience.iop.org/article/10.1088/0022-3727/28/11/003/meta}
\BIBentrySTDinterwordspacing

\bibitem{angiulli_characteristic_1998}
\BIBentryALTinterwordspacing
G.~Angiulli, G.~Amendola, and G.~D. Massa, ``Characteristic {Modes} in
  {Multiple} {Scattering} by {Conducting} {Cylinders} of {Arbitrary} {Shape},''
  \emph{Electromagnetics}, vol.~18, no.~6, pp. 593--612, Nov. 1998, publisher:
  Taylor \& Francis \_eprint: https://doi.org/10.1080/02726349808908615.
  [Online]. Available: \url{https://doi.org/10.1080/02726349808908615}
\BIBentrySTDinterwordspacing

\bibitem{doicu_light_2006}
\BIBentryALTinterwordspacing
A.~Doicu, T.~Wriedt, and Y.~A. Eremin, \emph{Light {Scattering} by {Systems} of
  {Particles}: {Null}-{Field} {Method} with {Discrete} {Sources}: {Theory} and
  {Programs}}, ser. Springer {Series} in {Optical} {Sciences}.\hskip 1em plus
  0.5em minus 0.4em\relax Berlin Heidelberg: Springer-Verlag, 2006. [Online].
  Available: \url{https://www.springer.com/gp/book/9783540336969}
\BIBentrySTDinterwordspacing

\bibitem{rao_electromagnetic_1982}
S.~Rao, D.~Wilton, and A.~Glisson, ``Electromagnetic scattering by surfaces of
  arbitrary shape,'' \emph{IEEE Transactions on Antennas and Propagation},
  vol.~30, no.~3, pp. 409--418, May 1982, conference Name: IEEE Transactions on
  Antennas and Propagation.

\bibitem{bossavit_computational_1998}
A.~Bossavit, \emph{Computational electromagnetism: variational formulations,
  complementarity, edge elements}.\hskip 1em plus 0.5em minus 0.4em\relax
  Academic Press, 1998.

\bibitem{albanese_integral_1988}
R.~Albanese and G.~Rubinacci, ``Integral formulation for {3D} eddy-current
  computation using edge elements,'' \emph{IEE Proceedings A (Physical Science,
  Measurement and Instrumentation, Management and Education, Reviews)}, vol.
  135, no.~7, pp. 457--462, 1988.

\bibitem{rubinacci_broadband_2006}
G.~Rubinacci and A.~Tamburrino, ``A broadband volume integral formulation based
  on edge-elements for full-wave analysis of lossy interconnects,'' \emph{IEEE
  transactions on antennas and propagation}, vol.~54, no.~10, pp. 2977--2989,
  2006.

\bibitem{wilton_novel_1993}
D.~Wilton, J.~Lim, and S.~Rao, ``A novel technique to calculate the
  electromagnetic scattering by surfaces of arbitrary shape,'' \emph{URSI Radio
  Science Meeting}, p. 322, 1993.

\bibitem{wu_study_1995}
W.-L. Wu, A.~W. Glisson, and D.~Kajfez, ``A study of two numerical solution
  procedures for the electric field integral equation at low frequency,''
  \emph{Applied Computational Electromagnetics Society Journal}, vol.~10,
  no.~3, pp. 69--80, 1995.

\bibitem{trintinalia_first_2001}
\BIBentryALTinterwordspacing
L.~Trintinalia and H.~Ling, ``First {Order} {Triangular} {Patch} {Basis}
  {Functions} for {Electromagnetic} {Scattering} {Analysis},'' \emph{Journal of
  Electromagnetic Waves and Applications}, vol.~15, no.~11, pp. 1521--1537,
  Jan. 2001, publisher: Taylor \& Francis \_eprint:
  https://doi.org/10.1163/156939301X00085. [Online]. Available:
  \url{https://doi.org/10.1163/156939301X00085}
\BIBentrySTDinterwordspacing

\bibitem{buffa_dual_2007}
\BIBentryALTinterwordspacing
A.~Buffa and S.~H. Christiansen, ``A {Dual} {Finite} {Element} {Complex} on the
  {Barycentric} {Refinement},'' \emph{Mathematics of Computation}, vol.~76, no.
  260, pp. 1743--1769, 2007, publisher: American Mathematical Society.
  [Online]. Available: \url{https://www.jstor.org/stable/40234460}
\BIBentrySTDinterwordspacing

\bibitem{graglia_higher_1997}
R.~Graglia, D.~Wilton, and A.~Peterson, ``Higher order interpolatory vector
  bases for computational electromagnetics,'' \emph{IEEE Transactions on
  Antennas and Propagation}, vol.~45, no.~3, pp. 329--342, Mar. 1997,
  conference Name: IEEE Transactions on Antennas and Propagation.

\bibitem{bohren_absorption_1998}
C.~F. Bohren and D.~R. Huffman, \emph{Absorption and {Scattering} of {Light} by
  {Small} {Particles}}.\hskip 1em plus 0.5em minus 0.4em\relax Wiley, 1998.

\bibitem{li_spheroidal_2004}
L.-W. Li, X.-K. Kang, and M.-S. Leong, \emph{\BIBforeignlanguage{en}{Spheroidal
  {Wave} {Functions} in {Electromagnetic} {Theory}}}.\hskip 1em plus 0.5em
  minus 0.4em\relax John Wiley \& Sons, Apr. 2004.

\bibitem{garbacz_modal_1965}
R.~Garbacz, ``Modal expansions for resonance scattering phenomena,''
  \emph{Proceedings of the IEEE}, vol.~53, no.~8, pp. 856--864, 1965.

\bibitem{chang_surface_1977}
Y.~Chang and R.~Harrington, ``A surface formulation for characteristic modes of
  material bodies,'' \emph{IEEE Transactions on Antennas and Propagation},
  vol.~25, no.~6, pp. 789--795, Nov. 1977, conference Name: IEEE Transactions
  on Antennas and Propagation.

\bibitem{harrington_characteristic_1972}
R.~Harrington, J.~Mautz, and {Yu Chang}, ``Characteristic modes for dielectric
  and magnetic bodies,'' \emph{IEEE Transactions on Antennas and Propagation},
  vol.~20, no.~2, pp. 194--198, Mar. 1972.

\bibitem{chen_characteristic_2015}
Y.~Chen and C.-F. Wang, \emph{\BIBforeignlanguage{en}{Characteristic {Modes}:
  {Theory} and {Applications} in {Antenna} {Engineering}}}.\hskip 1em plus
  0.5em minus 0.4em\relax John Wiley \& Sons, Jun. 2015, google-Books-ID:
  suobBgAAQBAJ.

\bibitem{faenzi_metasurface_2019}
\BIBentryALTinterwordspacing
M.~Faenzi, G.~Minatti, D.~González-Ovejero, F.~Caminita, E.~Martini,
  C.~Della~Giovampaola, and S.~Maci, ``Metasurface {Antennas}: {New} {Models},
  {Applications} and {Realizations},'' \emph{Scientific Reports}, vol.~9,
  no.~1, p. 10178, Jul. 2019. [Online]. Available:
  \url{https://www.nature.com/articles/s41598-019-46522-z}
\BIBentrySTDinterwordspacing

\bibitem{forestiere_electromagnetic_2019}
\BIBentryALTinterwordspacing
C.~Forestiere, G.~Gravina, G.~Miano, M.~Pascale, and R.~Tricarico,
  ``Electromagnetic modes and resonances of two-dimensional bodies,''
  \emph{Physical Review B}, vol.~99, no.~15, p. 155423, Apr. 2019, publisher:
  American Physical Society. [Online]. Available:
  \url{https://link.aps.org/doi/10.1103/PhysRevB.99.155423}
\BIBentrySTDinterwordspacing

\bibitem{mayergoyz_electrostatic_2005}
\BIBentryALTinterwordspacing
I.~D. Mayergoyz, D.~R. Fredkin, and Z.~Zhang, ``Electrostatic (plasmon)
  resonances in nanoparticles,'' \emph{Phys. Rev. B}, vol.~72, no.~15, p.
  155412, Oct. 2005. [Online]. Available:
  \url{https://link.aps.org/doi/10.1103/PhysRevB.72.155412}
\BIBentrySTDinterwordspacing

\bibitem{miano_numerical_2010}
G.~Miano, G.~Rubinacci, and A.~Tamburrino, ``Numerical {Modeling} for the
  {Analysis} of {Plasmon} {Oscillations} in {Metallic} {Nanoparticles},''
  \emph{IEEE Transactions on Antennas and Propagation}, vol.~58, no.~9, pp.
  2920--2933, 2010.

\bibitem{forestiere_magnetoquasistatic_2020}
\BIBentryALTinterwordspacing
C.~Forestiere, G.~Miano, G.~Rubinacci, M.~Pascale, A.~Tamburrino, R.~Tricarico,
  and S.~Ventre, ``Magnetoquasistatic resonances of small dielectric objects,''
  \emph{Phys. Rev. Research}, vol.~2, no.~1, p. 013158, Feb. 2020. [Online].
  Available: \url{https://link.aps.org/doi/10.1103/PhysRevResearch.2.013158}
\BIBentrySTDinterwordspacing

\bibitem{forestiere_quantum_2020}
\BIBentryALTinterwordspacing
C.~Forestiere, G.~Miano, M.~Pascale, and R.~Tricarico, ``Quantum theory of
  radiative decay rate and frequency shift of surface plasmon modes,''
  \emph{Physical Review A}, vol. 102, no.~4, p. 043704, Oct. 2020, publisher:
  American Physical Society. [Online]. Available:
  \url{https://link.aps.org/doi/10.1103/PhysRevA.102.043704}
\BIBentrySTDinterwordspacing

\bibitem{forestiere_operative_2022}
\BIBentryALTinterwordspacing
C.~Forestiere and G.~Miano, ``Operative approach to quantum electrodynamics in
  dispersive dielectric objects based on a polarization-mode expansion,''
  \emph{Physical Review A}, vol. 106, no.~3, p. 033701, Sep. 2022, publisher:
  American Physical Society. [Online]. Available:
  \url{https://link.aps.org/doi/10.1103/PhysRevA.106.033701}
\BIBentrySTDinterwordspacing

\bibitem{forestiere_resonance_2020}
\BIBentryALTinterwordspacing
C.~Forestiere, G.~Miano, and G.~Rubinacci, ``Resonance frequency and radiative
  {Q}-factor of plasmonic and dieletric modes of small objects,'' \emph{Phys.
  Rev. Research}, vol.~2, no.~4, p. 043176, Nov. 2020. [Online]. Available:
  \url{https://link.aps.org/doi/10.1103/PhysRevResearch.2.043176}
\BIBentrySTDinterwordspacing

\bibitem{forestiere_time-domain_2021}
\BIBentryALTinterwordspacing
C.~Forestiere and G.~Miano, ``Time-domain formulation of electromagnetic
  scattering based on a polarization-mode expansion and the principle of least
  action,'' \emph{Physical Review A}, vol. 104, no.~1, p. 013512, Jul. 2021,
  publisher: American Physical Society. [Online]. Available:
  \url{https://link.aps.org/doi/10.1103/PhysRevA.104.013512}
\BIBentrySTDinterwordspacing

\bibitem{suter_subdomain_2000}
\BIBentryALTinterwordspacing
E.~Suter and J.~R. Mosig, ``\BIBforeignlanguage{en}{A subdomain multilevel
  approach for the efficient {MoM} analysis of large planar antennas},''
  \emph{\BIBforeignlanguage{en}{Microwave and Optical Technology Letters}},
  vol.~26, no.~4, pp. 270--277, 2000
\BIBentrySTDinterwordspacing

\bibitem{prakash_characteristic_2003}
\BIBentryALTinterwordspacing
V.~V.~S. Prakash and R.~Mittra, ``\BIBforeignlanguage{en}{Characteristic basis
  function method: {A} new technique for efficient solution of method of
  moments matrix equations},'' \emph{\BIBforeignlanguage{en}{Microwave and
  Optical Technology Letters}}, vol.~36, no.~2, pp. 95--100, 2003, \_eprint:
  https://onlinelibrary.wiley.com/doi/pdf/10.1002/mop.10685. [Online].
  Available: \url{https://onlinelibrary.wiley.com/doi/abs/10.1002/mop.10685}
\BIBentrySTDinterwordspacing

\bibitem{matekovits_analysis_2007}
L.~Matekovits, V.~A. Laza, and G.~Vecchi, ``Analysis of {Large} {Complex}
  {Structures} {With} the {Synthetic}-{Functions} {Approach},'' \emph{IEEE
  Transactions on Antennas and Propagation}, vol.~55, no.~9, pp. 2509--2521,
  Sep. 2007, conference Name: IEEE Transactions on Antennas and Propagation.

\bibitem{freni_fast-factorization_2011}
A.~Freni, P.~De~Vita, P.~Pirinoli, L.~Matekovits, and G.~Vecchi,
  ``Fast-{Factorization} {Acceleration} of {MoM} {Compressive}
  {Domain}-{Decomposition},'' \emph{IEEE Transactions on Antennas and
  Propagation}, vol.~59, no.~12, pp. 4588--4599, Dec. 2011, conference Name:
  IEEE Transactions on Antennas and Propagation.

\bibitem{vecchi_hybrid_1996}
G.~Vecchi, L.~Matekovits, P.~Pirinoli, and M.~Orefice, ``Hybrid
  spectral-spatial method for the analysis of printed antennas,'' \emph{Radio
  Science}, vol.~31, no.~5, pp. 1263--1270, Sep. 1996, conference Name: Radio
  Science.

\bibitem{vecchi_numerical_1997}
\BIBentryALTinterwordspacing
------, ``\BIBforeignlanguage{en}{A numerical regularization of the {EFIE} for
  three-dimensional planar structures in layered media (invited article)},''
  \emph{\BIBforeignlanguage{en}{International Journal of Microwave and
  Millimeter-Wave Computer-Aided Engineering}}, vol.~7, no.~6, pp. 410--431,
  1997
\BIBentrySTDinterwordspacing

\bibitem{wu_scattering_1977}
\BIBentryALTinterwordspacing
T.-K. Wu and L.~L. Tsai, ``\BIBforeignlanguage{en}{Scattering from
  arbitrarily-shaped lossy dielectric bodies of revolution},''
  \emph{\BIBforeignlanguage{en}{Radio Science}}, vol.~12, no.~5, pp. 709--718,
  1977
\BIBentrySTDinterwordspacing

\bibitem{poggio_chapter_1973}
\BIBentryALTinterwordspacing
A.~J. Poggio and E.~K. Miller, ``\BIBforeignlanguage{en}{{CHAPTER} 4 -
  {Integral} {Equation} {Solutions} of {Three}-dimensional {Scattering}
  {Problems}},'' in \emph{\BIBforeignlanguage{en}{Computer {Techniques} for
  {Electromagnetics}}}, ser. International {Series} of {Monographs} in
  {Electrical} {Engineering}, R.~Mittra, Ed.\hskip 1em plus 0.5em minus
  0.4em\relax Pergamon, Jan. 1973, pp. 159--264.
\BIBentrySTDinterwordspacing

\bibitem{harrington_field_1993}
R.~F. Harrington, \emph{Field computation by moment methods}.\hskip 1em plus
  0.5em minus 0.4em\relax Wiley-IEEE Press, 1993.

\bibitem{monk_finite_2003}
P.~Monk and D.~o. M. S. P.~M. PH, \emph{\BIBforeignlanguage{en}{Finite
  {Element} {Methods} for {Maxwell}'s {Equations}}}.\hskip 1em plus 0.5em minus
  0.4em\relax Clarendon Press, Apr. 2003.

\bibitem{scharstein_helmholtz_1991}
R.~Scharstein, ``Helmholtz decomposition of surface electric current in
  electromagnetic scattering problems,'' in \emph{[1991 {Proceedings}] {The}
  {Twenty}-{Third} {Southeastern} {Symposium} on {System} {Theory}}, Mar. 1991,
  pp. 424--426, iSSN: 0094-2898.

\bibitem{nair_generalized_2011}
N.~V. Nair and B.~Shanker, ``Generalized {Method} of {Moments}: {A} {Novel}
  {Discretization} {Technique} for {Integral} {Equations},'' \emph{IEEE
  Transactions on Antennas and Propagation}, vol.~59, no.~6, pp. 2280--2293,
  Jun. 2011, conference Name: IEEE Transactions on Antennas and Propagation.

\bibitem{tamburrino_monotonicity_2021}
\BIBentryALTinterwordspacing
A.~Tamburrino, G.~Piscitelli, and Z.~Zhou, ``\BIBforeignlanguage{en}{The
  monotonicity principle for magnetic induction tomography},''
  \emph{\BIBforeignlanguage{en}{Inverse Problems}}, vol.~37, no.~9, p. 095003,
  Aug. 2021.
\BIBentrySTDinterwordspacing

\bibitem{burton_study_1995}
M.~Burton and S.~Kashyap, ``A study of a recent, moment-method algorithm that
  is accurate to very low frequencies,'' \emph{Applied Computational
  Electromagnetics Society Journal}, vol.~10, pp. 58--68, 1995, publisher:
  APPLIED COMPUTATIONAL ELECTROMAGNETICS SOCIETY INC.

\bibitem{golub_matrix_1983}
H.~Golub and C.~F.~V. Loan, \emph{Matrix {Computations}}.\hskip 1em plus 0.5em
  minus 0.4em\relax Baltimore, MD: Johns Hopkins University Press, 1983.

\bibitem{graglia_numerical_1993}
R.~Graglia, ``On the numerical integration of the linear shape functions times
  the 3-{D} {Green}'s function or its gradient on a plane triangle,''
  \emph{Antennas and Propagation, IEEE Transactions on}, vol.~41, no.~10, pp.
  1448 --1455, Oct. 1993.

\bibitem{ubeda_divergence-conforming_2011}
E.~Ubeda and J.~M. Rius, ``Divergence-conforming discretization of second-kind
  integral equations for the {RCS} computation in the {Rayleigh} frequency
  region,'' \emph{Radio Science}, vol.~46, no.~05, pp. 1--10, Oct. 2011,
  conference Name: Radio Science.

\bibitem{wilton_improving_1981}
D.~R. Wilton and A.~W. Glisson, ``On improving the electric field inte-gral
  equation at low frequencies,'' Los Angeles, CA, Jun. 1981, pp. 22--24.

\bibitem{vecchi_loop-star_1999}
G.~Vecchi, ``Loop-star decomposition of basis functions in the discretization
  of the {EFIE},'' \emph{IEEE Transactions on Antennas and Propagation},
  vol.~47, no.~2, pp. 339--346, Feb. 1999, conference Name: IEEE Transactions
  on Antennas and Propagation.

\bibitem{andriulli_loop-star_2012}
F.~P. Andriulli, ``Loop-{Star} and {Loop}-{Tree} {Decompositions}: {Analysis}
  and {Efficient} {Algorithms},'' \emph{IEEE Transactions on Antennas and
  Propagation}, vol.~60, no.~5, pp. 2347--2356, May 2012, conference Name: IEEE
  Transactions on Antennas and Propagation.

\bibitem{miano_surface_2005}
G.~Miano and F.~Villone, ``A surface integral formulation of {Maxwell}
  equations for topologically complex conducting domains,'' \emph{IEEE
  Transactions on Antennas and Propagation}, vol.~53, no.~12, pp. 4001--4014,
  Dec. 2005, conference Name: IEEE Transactions on Antennas and Propagation.

\bibitem{zhao_integral_2000}
J.-S. Zhao and W.~C. Chew, ``Integral equation solution of {Maxwell}'s
  equations from zero frequency to microwave frequencies,'' \emph{IEEE
  Transactions on Antennas and Propagation}, vol.~48, no.~10, pp. 1635--1645,
  Oct. 2000, conference Name: IEEE Transactions on Antennas and Propagation.

\bibitem{chen_analysis_2001}
S.~Chen, W.~C. Chew, J.~Song, and J.-S. Zhao, ``Analysis of low frequency
  scattering from penetrable scatterers,'' \emph{IEEE Transactions on
  Geoscience and Remote Sensing}, vol.~39, no.~4, pp. 726--735, Apr. 2001,
  conference Name: IEEE Transactions on Geoscience and Remote Sensing.

\bibitem{yla-oijala_surface_2018}
P.~Ylä-Oijala, H.~Wallén, D.~C. Tzarouchis, and A.~Sihvola, ``Surface
  {Integral} {Equation}-{Based} {Characteristic} {Mode} {Formulation} for
  {Penetrable} {Bodies},'' \emph{IEEE Transactions on Antennas and
  Propagation}, vol.~66, no.~7, pp. 3532--3539, Jul. 2018, conference Name:
  IEEE Transactions on Antennas and Propagation.

\bibitem{taskinen_current_2006}
M.~Taskinen and P.~Yla-Oijala, ``Current and charge {Integral} equation
  formulation,'' \emph{IEEE Transactions on Antennas and Propagation}, vol.~54,
  no.~1, pp. 58--67, Jan. 2006, conference Name: IEEE Transactions on Antennas
  and Propagation.

\bibitem{qian_augmented_2008}
\BIBentryALTinterwordspacing
Z.~G. Qian and W.~C. Chew, ``\BIBforeignlanguage{en}{An augmented electric
  field integral equation for high-speed interconnect analysis},''
  \emph{\BIBforeignlanguage{en}{Microwave and Optical Technology Letters}},
  vol.~50, no.~10, pp. 2658--2662, 2008, \_eprint:
  https://onlinelibrary.wiley.com/doi/pdf/10.1002/mop.23736. [Online].
  Available: \url{https://onlinelibrary.wiley.com/doi/abs/10.1002/mop.23736}
\BIBentrySTDinterwordspacing

\bibitem{forestiere_frequency_2017}
C.~Forestiere, G.~Miano, G.~Rubinacci, A.~Tamburrino, L.~Udpa, and S.~Ventre,
  ``A {Frequency} {Stable} {Volume} {Integral} {Equation} {Method} for
  {Anisotropic} {Scatterers},'' \emph{IEEE Transactions on Antennas and
  Propagation}, vol.~65, no.~3, pp. 1224--1235, Mar. 2017.

\bibitem{mie_beitrage_1908}
G.~Mie, ``Beiträge zur {Optik} trüber {Medien}, speziell kolloidaler
  {Metallösungen},'' \emph{Annalen der physik}, vol. 330, no.~3, pp. 377--445,
  1908.

\bibitem{johnson_optical_1972}
\BIBentryALTinterwordspacing
P.~B. Johnson and R.~W. Christy, ``Optical {Constants} of the {Noble}
  {Metals},'' \emph{Phys. Rev. B}, vol.~6, no.~12, pp. 4370--4379, Dec. 1972.
  [Online]. Available: \url{https://link.aps.org/doi/10.1103/PhysRevB.6.4370}
\BIBentrySTDinterwordspacing

\bibitem{schuller_plasmonics_2010}
J.~A. Schuller, E.~S. Barnard, W.~Cai, Y.~C. Jun, J.~S. White, and M.~L.
  Brongersma, ``Plasmonics for extreme light concentration and manipulation,''
  \emph{Nat. Mater.}, vol.~9, pp. 193--204, 2010.

\bibitem{persson_simple_2004}
\BIBentryALTinterwordspacing
P.-O. Persson and G.~Strang, ``A {Simple} {Mesh} {Generator} in {Matlab},''
  \emph{SIAM Review}, vol.~46, no.~2, pp. 329--345, 2004, publisher: Society
  for Industrial and Applied Mathematics. [Online]. Available:
  \url{https://www.jstor.org/stable/20453511}
\BIBentrySTDinterwordspacing

\bibitem{nosich_method_1999}
A.~Nosich, ``The method of analytical regularization in wave-scattering and
  eigenvalue problems: foundations and review of solutions,'' \emph{IEEE
  Antennas and Propagation Magazine}, vol.~41, no.~3, pp. 34--49, Jun. 1999,
  conference Name: IEEE Antennas and Propagation Magazine.

\end{thebibliography}
\end{document}